\documentclass[preprint, 3p]{elsarticle}

\makeatletter
\def\ps@pprintTitle{%
  \let\@oddhead\@empty
  \let\@evenhead\@empty
  \def\@oddfoot{\normalfont\hfil\thepage\hfil}
  \let\@evenfoot\@oddfoot
}
\makeatother

\usepackage{preamble}

\begin{document}

\journal{}

\begin{frontmatter}

    \title{Niching Importance Sampling for Multi-modal Rare-event Simulation}

\author[ucl]{Hugh J. Kinnear}
\author[ucl]{F.A. DiazDelaO}

\affiliation[ucl]{organization={Clinical Operational Research Unit, Department of Mathematics},
            addressline={ UCL},
            city={London},
            postcode={WC1H 0BT},
            country={United Kingdom}}

\begin{abstract}

This paper proposes niching importance sampling, a framework that combines concepts from reliability analysis, e.g. Markov chains, importance sampling, and relative cross entropy minimisation, with niching techniques from evolutionary multi-modal optimisation. The result is a highly robust estimator of the probability of failure, that can tackle sampling challenges posed by the underlying geometry of a reliability problem. Niching importance sampling is tested on a range of numerical examples and is shown to consistently avoid the degenerate behaviour observed for existing reliability methods on several multi-modal performance functions.

\end{abstract}


\end{frontmatter}

\section*{Acronyms}

\begin{tabular}{p{0.09\textwidth}p{0.36\textwidth}p{0.11\textwidth}p{0.34\textwidth}}
    \textbf{\acs{MC}} & Monte Carlo & \textbf{\acs{CE}} & Cross entropy \\
    \textbf{\acs{IS}} & Importance sampling & \textbf{\acs{NIS}} & Niching importance sampling \\
    \textbf{\acs{DIS}} & Direct importance sampling & \textbf{\acs{NInitS}} & Niching initial sampling \\
    \textbf{\acs{SuS}} & Subset simulation & \textbf{\acs{vMF}} & Von Mises-Fisher \\
    \textbf{\acs{SIS}} & Sequential importance sampling & \textbf{\acs{vMFNM}} & Von Mises-Fisher-Nakagami mixture \\
    \textbf{\acs{SAIS}} & Sequential adaptive imp. samp. & \textbf{\acs{EM}} & Expectation-maximisation 
\end{tabular}

\section{Introduction}\label{sec:introduction}

In reliability analysis, the behaviour of a physical system is modelled by a performance function operating on a space of uncertain input variables. The probability of failure, the primary quantity of interest in this context, is defined as the probability measure of the failure region, which refers to the set of input configurations for which the demand exceeds the system's capacity. Over the past few decades, various reliability methods have been developed to estimate the probability of failure \cite{songMonteCarloVariance2023}. The suitability of any of these reliability methods is determined by the characteristics of the particular reliability problem being considered. When the Taylor expansion of the performance function around appropriate points in the input space yields a good approximation, first- and second-order reliability methods can provide an accurate estimate of the probability of failure with low computational cost \cite{amExactInvariantFirst1974,keshtegarHybridSelfadaptiveConjugate2018,derkiureghianSecondOrderReliability1987,huangNewDirectSecondorder2018}. When the performance function is computationally expensive to evaluate, a surrogate model can be built and analysed \cite{bucherFastEfficientResponse1990,kaymazApplicationKrigingMethod2005,bourinetRareeventProbabilityEstimation2016,cardosoStructuralReliabilityAnalysis2008}. Alternatively, if little is known about a reliability problem, \ac{MC}-based methods \cite{schuellerEfficientMonteCarlo2009} offer a robust approach for estimating the probability of failure.

When a physical system is well-designed, the probability of failure tends to be, by definition, very small. Under these circumstances, a simple \ac{MC} estimator is highly inefficient in producing failure samples, that is, input configurations in the failure region. Variance reduction techniques are a class of reliability methods that seek to address this problem by modifying the simple \ac{MC} estimator. Here, \Ac{IS} is the fundamental concept which, explicitly or implicitly, underlies the theory of most variance reduction techniques \cite{tabandehReviewAssessmentImportance2022a}. The idea is to sample from a so-called importance distribution, which assigns high probability density to the most important regions of the input space. An \ac{IS} method is characterised by the strategy it employs to construct the importance distribution, since the computational efficiency of such methods depends entirely on how accurately the importance distribution models the failure region.

In this work, it will be useful to classify \ac{IS} methods for reliability analysis into two broad categories. The first type, which will be referred to as \ac{DIS} methods, consists of separate initial sampling and distribution fitting procedures. The simplest of the \ac{DIS} methods, and first to be developed, are the static methods \cite{melchersImportanceSamplingStructural1989}, where the initial sampling procedure generates failure samples which are then used to fit the parameters of an importance distribution. Adaptive methods begin in the same way as static methods. However, once the first importance distribution has been fitted, it is then sampled from, and those samples are used to fit the parameters of a new importance distribution. This process can be iterated many times and often results in a relatively efficient estimator \cite{bucherAdaptiveSamplingIterative1988,melchersSearchbasedImportanceSampling1990}. The second type of \ac{IS} method for reliability analysis runs the initial sampling and adaptive fitting procedures concurrently \cite{auEstimationSmallFailure2001,papaioannouSequentialImportanceSampling2016,papaioannouImprovedCrossEntropybased2019}. These more recently proposed algorithms will be referred to as \ac{SAIS} methods.  \ac{SAIS} methods do not attempt to model the failure region directly, and only yield failure samples at the end of the procedure. Such algorithms are appealing, since they have a relatively simple control flow and assume very little about the performance function. Indeed, they are often applied when the performance function is only known implicitly.

However, it is known that if the performance function exhibits challenging topology, such as rapidly changing output and multiple local optima, \ac{SAIS} methods can give unreliable estimates for the probability of failure \cite{breitungGeometryLimitState2019,breitungReturnDesignPoints2024}. From this point of view, \ac{SAIS} methods can be compared to stochastic optimisation algorithms. In the same way that optimisation algorithms may get stuck in local optima, \ac{SAIS} methods can be led away from important regions of the input space. This problem has prompted the development of more robust variance reduction methods. In stochastic spectral embedding-based reliability, the important regions of the input space are highlighted via a sequential partitioning procedure, and a surrogate is fit in each set of the partition \cite{wagnerRareEventEstimation2022}. Sequential space conversion scales the input distribution and replaces the performance function with a series of control variates, allowing global information about the performance function to be incorporated into the probability of failure estimate \cite{rashkiSESCNewSubset2021}.

Evolutionary algorithms are a class of biologically inspired optimisation meta-heuristics that can be applied to find multiple local optima of an objective function in multi-modal optimisation problems \cite{backEvolutionaryAlgorithmsParameter2023}. Niching techniques, where ``niching'' in biology refers to the role a specific species plays in an environment, are used within evolutionary algorithms to maintain samples in the neighbourhoods of multiple local optima, called niches \cite{liSeekingMultipleSolutions2017}. Due to the similarities between some variance reduction techniques and evolutionary algorithms, it is natural to suggest that niching techniques can be applied to reliability problems. For instance, the approach taken in \cite{derkiureghianMultipleDesignPoints1998} of adding a bulge to the performance function could be considered a niching technique. More explicitly, niching techniques have been combined with \ac{SuS} to produce a robust estimator for the probability of failure \cite{kinnearNichingSubsetSimulation2025a}.

This paper proposes \ac{NIS}, a reliability analysis strategy that integrates niching techniques into an importance sampling framework. \ac{NIS} is a \ac{DIS} method that attempts to maintain the wide applicability of \ac{SAIS} methods, whilst more carefully and deliberately exploring the failure region. The fundamental component of \ac{NIS} is \ac{NInitS}, which utilises niching techniques in order to efficiently populate all the important subsets of the failure region. The output of \ac{NInitS} can be used to fit a mixture importance distribution using an \ac{EM} algorithm without the need for an initialisation procedure. Performance functions with challenging topology can lead to poor ergodic behaviour of the Markov chains used to generate failure samples, and so a component weight correction routine is also carried out after the \ac{EM} algorithm. Finally, a heuristic for determining an appropriate computational budget is introduced, based on the mutual information of the mixture importance distribution.

This paper is organised as follows. Section \ref{sec:importance sampling} formally defines the probability of failure and gives an overview of \ac{IS} reliability methods. \ac{NInitS} is introduced in Section \ref{sec:niching initial sampling} and applied within \ac{NIS} in Section \ref{sec:niching importance sampling}. In Section \ref{sec:numerical examples}, illustrative high-dimensional numerical examples are presented, alongside more practical reliability problems. The paper is summarised and concluded in Section \ref{sec:conclusion}, including ideas for future directions of research.

\ifSubfilesClassLoaded{%
    \newpage
    \bibliography{references}%
}

\end{document}

\section{Importance Sampling (IS)}\label{sec:importance sampling}

\subsection{Reliability Analysis}\label{sec:reliability analysis}

The behaviour of a system in reliability analysis is modelled by a performance function, $g:\mathcal{X} \subseteq \R^d \rightarrow \R$, which assigns a scalar value to every combination of inputs in the input space, $\bm{x} \in \mathcal{X}$. The input space is endowed with a probability density function $f(\bm{x})$, called the input distribution. If the performance of a set of inputs exceeds a critical threshold $b$, then that set of inputs is said to be in the failure region, denoted as $F = \{\bm{x}\in\mathcal{X}\ : g(\bm{x})\geq b\}$. Without loss of generality, the critical threshold can be set to $0$. The central task of reliability analysis is to determine the probability of failure, namely,
\begin{equation}\label{eq:probability of failure}
    P_{F} := \int_{\mathcal{X}} \one_{F}(\bm{x})f(\bm{x})d\bm{x}
    = \E_f \left[ \one_{F}(\bm{x})\right].
\end{equation}
where $\one_{F}(\cdot)$ is an indicator function.

Reliability methods can be categorised according to the assumptions about the reliability problem they rely on. Typically, the more simplifying assumptions a reliability method makes, the more computationally efficient it is. However, this comes at the cost of reduced robustness and applicability. \ac{NIS}, the method proposed in this work, is intended to be more robust than \ac{SAIS} techniques, whilst incurring the least possible additional computational cost. The following are the properties of the reliability problem being considered:

\begin{itemize}
    \item The performance function is computationally expensive. The assumption is that the cost of any other procedure is negligible compared to the cost of evaluating the performance function. Thus, these evaluations will be used to measure computational complexity and should be minimised.

    \item The input space is high-dimensional. It is assumed that the performance function is sensitive to all inputs across the input space. This property precludes the use of dimension reduction techniques to simplify the reliability problem.

    \item The performance function is treated as a black-box model. Consequently, no analytical techniques or gradient information are assumed to be available. Also, no prior expert information regarding the performance function is available. This includes knowledge of the number of components in the reliability problem and areas of the input space guaranteed to be in the failure region for the purpose of initial sampling.

    \item The performance function has challenging topology. In this work, ``challenging topology'' refers to geometric features of the performance function or failure region, such as multiple important niches, local optima, and rapidly varying performance values, that make it difficult for sequential adaptive importance sampling methods to reliably populate all important regions.

    \item Augmenting the standard deviation of the input distribution is not guaranteed to increase the probability of failure. The opposite was assumed in \cite{rashkiSESCNewSubset2021}.

    \item The input distribution is the standard normal multivariate distribution.
\end{itemize}

The idea behind the last property is that for some general input distribution, it is often possible to apply suitable transformations to obtain an equivalent reliability problem in standard normal space. If the input distribution is explicitly known, the Rosenblatt transformation can be applied \cite{rosenblattRemarksMultivariateTransformation1952}. If only the marginal distributions and correlations are given, the joint probability distribution can be approximated by a Nataf distribution \cite{natafDeterminationDistributionsDont1962}. Whilst this assumption limits applicability, it is made here for ease of exposition, concreteness, and to isolate the main algorithmic contribution. Provided that an appropriate parametric family of probability densities and Markov chain algorithm can be identified, the approach in this paper could be applied to a reliability problem with an arbitrary input distribution. Such extensions are not investigated in this work, and complications associated with the chosen transformation, any induced dependence structure, and the resulting sampling scheme are left for future study.

The most robust reliability method, which makes no assumptions about the reliability problem, is a simple \ac{MC} estimator,
\begin{equation}\label{eq:dmc}
    \hat{P}_{\text{MC}} = \frac{1}{N} \sum_{i=1}^{N} \one_{F}(\bm{x}_i),
\end{equation}
where $\bm{x}_1,\dots,\bm{x}_N \sim f $. The problem with the \ac{MC} estimator is made clear by its \ac{CoV},
\begin{equation}\label{eq:dmc_cov}
    \delta_{\text{MC}} = \sqrt{\frac{1-P_F}{NP_F}}.
\end{equation}
When the probability of failure is very small, the \ac{CoV} of the \ac{MC} estimator becomes very large for a fixed $N$. Thus, obtaining an acceptable \ac{CoV} requires a very large sample size, which requires many computationally expensive performance function evaluations. Variance reduction techniques are a type of reliability method that seek to reduce the variance of the \ac{MC} estimator. The fundamental concept underlying most variance reduction techniques, either explicitly or implicitly, is \ac{IS}.

\subsection{The Concept}\label{sec:the concept}

The idea of \ac{IS} is to introduce an importance distribution $q$ into the probability of failure expression,
\begin{equation}\label{eq:importance ra}
    P_{F} = \int_{\mathcal{X}} \one_{F}(\bm{x})\frac{f(\bm{x})}{q(\bm{x})}q(\bm{x})d\bm{x}
    = \E_{q} \left[ \one_{F}(\bm{x}) \frac{f(\bm{x})}{q(\bm{x})}\right].
\end{equation}
This way, the expectation is recast with respect to the importance distribution, and importance samples $\bm{x}_1,\dots,\bm{x}_N \sim q $, can be used to estimate the probability of failure,
\begin{equation}\label{eq:importance sampling}
    \hat{P}_{\text{IS}} = \frac{1}{N} \sum_{i=1}^{N} \one_{F}(\bm{x}_i) \frac{f(\bm{x}_i)}{q(\bm{x}_i)},
\end{equation}
where $f(\bm{x}_i)/q(\bm{x}_i)$ are referred to as the importance weights. The \ac{CoV} of the IS estimator is given by
\begin{equation}\label{eq:is cov}
    \delta_{\text{IS}} = \sqrt{\frac{\V_q \left[ \one_{F}(\bm{x}) \frac{f(\bm{x})}{q(\bm{x})}\right]}{N P_F^{2}}}.
\end{equation}
If the importance distribution is chosen appropriately, the \ac{CoV} of an IS estimator can be considerably smaller than the \ac{CoV} of an \ac{MC} estimator, given the same number of samples. The optimal importance distribution in reliability analysis, i.e., the one which minimises the variance of the IS estimator, is given by
\begin{equation}\label{eq:is optimal}
    q^{*}(\bm{x}) = \argmin_q \V \left[ \hat{P}_{IS} \right] =  \frac{\one_{F}(\bm{x})f(\bm{x})}{P_F}.
\end{equation}
An \ac{IS} estimator using $q^*$ has zero variance. However, $q^*$ is not useful in practice because the probability of failure appears in its definition. Additionally, it is generally not possible to sample efficiently from $q^*$ without the use of Markov chains.

Variance reduction methods apply \ac{IS} in varied ways. As mentioned in the introduction, one useful
classification of \ac{IS} techniques distinguishes methods that use an explicit initial sampling procedure (\ac{DIS} methods) from those that do not (\ac{SAIS} methods). Initial sampling generates failure samples and allows \ac{DIS} methods to target the optimal importance density directly. Instead, \ac{SAIS} methods target a sequence of intermediate distributions. \ac{SAIS} methods concurrently locate the failure region whilst fitting a sequence of importance distributions.
The sequence of intermediate distributions that \ac{SAIS} methods employ must be constructed carefully so that an efficient and accurate estimate of the probability of failure can be made. As the sequence progresses, the intermediate distributions must become better approximations of the optimal importance distribution. Additionally, it is crucial that consecutive intermediate densities are sufficiently close to one another, such that samples from one can be used in a procedure to efficiently produce samples from the next. The first intermediate distribution should be relatively easy to sample from, and is typically the input distribution itself. The sequences have the general form
\begin{equation}\label{eq:targ seq}
    q_i(\bm{x}) = \frac{\eta_i(\bm{x})}{P_i},
\end{equation}
for $0 \leq i \leq m$, where $P_i$ are the intermediate normalising constants and $\eta_i$ are the intermediate unnormalised densities. There are mainly two sequences that have been used as the numerator of Equation \ref{eq:targ seq} in the reliability analysis literature:
\begin{align}\label{eq:target seq options}
    \eta^1_i(\bm{x}) &= \one_{F_i}(\bm{x})f(\bm{x}),\\
    \eta^2_i(\bm{x}) &= \Phi \left(\frac{g(\bm{x})}{\sigma_i}\right) f(\bm{x}),
\end{align}
where $\Phi(\cdot)$ is the normal cdf and $F_i = \{ \bm{x} \in \mathcal{X}: g(\bm{x}) \geq b_i\}$ are intermediate failure regions with intermediate failure thresholds $b_i \leq 0$, and $\sigma_i > 0$ denote intermediate scale parameters. The sequences are constructed in such a way that $b_i \leq b_{i+1}$ and $\sigma_i \geq \sigma_{i+1}$ to ensure that, as the sequence progresses, the intermediate distributions become better approximations of the optimal importance density. Indeed, note that if $b_i = 0$ then $\one_{F_i}(\bm{x}) = \one_{F}(\bm{x})$ and that $\lim_{\sigma \rightarrow 0} \Phi \left(g(\bm{x})/\sigma\right) = \one_{F}(\bm{x})$. Typically, the intermediate failure thresholds and intermediate scale parameters are chosen adaptively as the algorithm progresses to guarantee that the consecutive intermediate distributions are sufficiently close to one another.

In the remainder of this section two \ac{SAIS} methods, \ac{CE} and \ac{SIS}, will be briefly described so that the problems with such variance reduction methods can be understood and so that some of the ideas can be used in Section \ref{sec:niching initial sampling} and Section \ref{sec:niching importance sampling}.

\subsection{Cross Entropy (CE)}\label{sec:cross entropy}

At the $i^\text{th}$ step, sequential adaptive \ac{CE} methods attempt to approximate the intermediate density $q_i$, with a parametric family of probability densities $h(\bm{x};\bm{\nu})$ where $\bm{\nu}$ denotes the vector of parameters. It should be possible to evaluate all the probability densities in the family point-wise, and they should be relatively easy to sample from. The \ac{KL} divergence, also known as the relative cross entropy, can be used to quantify how close two probability densities are,
\begin{equation}\label{eq:kl}
    D(q_i(\bm{x}),h(\bm{x};\bm{\nu})) = \E_{q_i}\left[\ln\left(\frac{q_i(\bm{x})}{h(\bm{x};\bm{\nu})}\right)\right].
\end{equation}
The goal in \ac{CE} is to find the parameters $\bm{\nu}_i$ that minimise the \ac{KL} divergence to the intermediate density,
\begin{align}
    \bm{\nu}_i &= \argmin_{\bm{\nu}} D(q_i(\bm{x}),h(\bm{x};\bm{\nu})) \label{eq:kl opt_1} \\
    & = \argmax_{\bm{\nu}} \E_{q_{i}}\left[\ln(h(\bm{x};\bm{\nu})) \right]. \label{eq:kl opt_2}
\end{align}

The expectation in Equation \ref{eq:kl opt_2} can be approximated with a \ac{MC} estimator using samples from $q_i$. However, it is not generally possible to sample efficiently from $q_i$ without a Markov chain. To overcome this challenge, the expectation can be rewritten with respect to the approximate density from the previous step, that is, $h(\bm{x};\bm{\nu}_{i-1})$:
\begin{equation}\label{eq:cross prev params}
    \bm{\nu}_{i} = \argmax_{\bm{\nu}} \E_{h_{\bm{\nu}_{i-1}}} \left[ \ln(h(\bm{x};\bm{\nu})) \frac{\eta_i(\bm{x})}{h(\bm{x};\bm{\nu}_{i-1})}\right].
\end{equation}
Note that the normalising constant $P_i$ does not appear in Equation \ref{eq:cross prev params} since it does not affect the optimisation problem. Samples $\bm{x}_1,\dots,\bm{x}_N \sim h(\cdot \, ;\bm{\nu}_{i-1}) $ can be used to estimate the expectation,
\begin{equation}\label{eq:approx cross}
    \hat{\bm{\nu}}_{i} = \argmax_{\bm{\nu}} \frac{1}{N} \sum_{j=1}^N \ln(h(\bm{x}_j;\bm{\nu})) \frac{\eta_i(\bm{x}_j)}{h(\bm{x}_j;\hat{\bm{\nu}}_{i-1})}.
\end{equation}
Updating rules for solving the optimisation problem, dependent on the parametric family chosen, can typically be derived by setting the gradient of the objective function equal to zero.

Originally, the unnormalised density sequence $\eta^1_i$ was used for sequential adaptive \ac{CE} \cite{kurtzCrossentropybasedAdaptiveImportance2013,geyerCrossEntropybasedImportance2019}. More recently however, a method referred to as \ac{iCE} was proposed which uses $\eta^2_i$ instead \cite{papaioannouImprovedCrossEntropybased2019}. It should be noted that \ac{CE} reliability methods have been suggested which employ an initial sampling procedure rather than an intermediate sequence \cite{wangCrossentropybasedAdaptiveImportance2016,mehniReliabilityAnalysisCrossentropy2023}. For more explicit details of \ac{CE} reliability methods, the reader is directed towards the references that have been used in this section.

\subsection{Sequential Importance Sampling (SIS)}\label{sec:sequential importance sampling}

In contrast to \ac{CE}, \ac{SIS} directly samples from the intermediate unnormalised densities, rather than some approximation. This is accomplished using Markov chains, which are able to sample without knowing normalisation constants. The distribution from which a particular Markov chain samples is known as its stationary distribution. Consecutive normalisation constants can be related as follows:
\begin{equation}\label{eq:conseq constants}
    P_i = P_{i-1}\int_{\mathcal{X}} \frac{\eta_i(\bm{x})}{\eta_{i-1}(\bm{x})}q_{i-1}(\bm{x})  d\bm{x} = P_{i-1}  \E_{q_{i-1}} \left[ \frac{\eta_i(\bm{x})}{\eta_{i-1}(\bm{x})}\right].
\end{equation}
By rearranging Equation \ref{eq:conseq constants}, and replacing the expectation with an estimator based on Markov chain samples $\bm{x}_1,\dots,\bm{x}_N$ with stationary distribution $q_{i-1}$, an estimator for the ratio of normalising constants results in
\begin{equation}\label{eq:ratio}
    \frac{P_i}{P_{i-1}} \approx \hat{S}_i   = \frac{1}{N} \sum_{j=1}^{N} \frac{\eta_i(\bm{x}_j)}{\eta_{i-1}(\bm{x}_{j})},
\end{equation}
for $1 \leq i \leq m$. If the intermediate densities are chosen in such a way that $P_0 = 1$ and $P_m = P_F$, then the \ac{SIS} estimator for the probability of failure is
\begin{equation}\label{eq:sis estimate}
    \hat{P}_{\text{SIS}}= \prod_{i=1}^{m} \hat{S}_i.
\end{equation}

To sample from $q_{i-1}$, the Markov chains use samples from the previous step as starting points, also known as seeds. It should be noted that this kind of sampling has some drawbacks. Markov chains typically require a burn-in period before they begin sampling from the stationary distribution. Also, the samples are correlated and so the resulting estimators usually require more samples for the same variance when compared to independent samples. When $\eta^1_i$ is used as intermediate density sequence, the resulting method is Subset Simulation \cite{auEstimationSmallFailure2001}. When $\eta^2_i$ is used, the resulting method is called  and \ac{SIS} \cite{papaioannouSequentialImportanceSampling2016}. For more explicit details of \ac{SIS} reliability methods, the reader is directed to the references cited in this section.

\subsection{Challenges and Motivating Examples}\label{sec:issues}

The limit state surface and the design point are two important concepts in reliability analysis. The former is defined as the set $\Lambda = \{\bm{x} \in \mathcal{X}: g(\bm{x}) = b \}$ and the latter is defined as the point $\bm{x} \in \Lambda$ closest to the origin, assuming that the input distribution is a standard multivariate normal. An equivalent definition is that it is the point in the failure region with the highest probability density. This means, particularly in the case of a low-dimensional input space, that the neighbourhood of the design point often contributes a significant amount to the probability of failure. It follows that any variance reduction technique that fails to populate neighbourhoods that contribute substantially to the probability of failure, including the neighbourhood of the design point when it is dominant, will tend to underestimate the probability of failure. In this work, a niche is regarded as important when its associated region contributes materially to the probability of failure, not merely because it contains a local optimum of the performance function.

As has been noted before \cite{breitungGeometryLimitState2019,breitungReturnDesignPoints2024}, sequential adaptive \ac{IS} methods like \ac{CE} and \ac{SIS} can be interpreted heuristically as performing a stochastic search toward higher-performance regions. As these algorithms progress, the intermediate distributions increasingly favour samples with higher performance, so the sample population tends to move toward regions that are closer to, or inside, the failure region. This behaviour is encouraged both by the intermediate densities themselves and by generating new samples in the neighbourhoods of currently high-performance samples. This type of greedy search is often a sensible heuristic for locating the failure region. However, performance functions with features such as rapidly changing output and multiple local optima, can lead \ac{SAIS} methods astray, ultimately resulting in no samples in the neighbourhood of the design point and an underestimation of the probability of failure. The difficulty is not that failure samples cannot be generated in principle for such examples, but rather that sequential adaptive methods such as \ac{SIS} may fail to discover and populate the relevant failure-region neighbourhoods reliably without additional structural information.

In order to understand the behaviour of \ac{SAIS} methods, it is useful to consider the following dynamical system, which can be defined for any differentiable performance function $g$,
\begin{equation}\label{eq:dynamical system}
    \frac{d \bm{x}(t)}{dt} = \nabla g(\bm{x}(t)),
\end{equation}
where $t$ is a time-like variable and $\bm{x}(t)$ is a solution trajectory \cite{klarbringDynamicalSystemsTopology2010}. For any point $\bm{x}_0$ in the input space, there exists an associated solution trajectory with initial condition $\bm{x}(0) = \bm{x}_0$ and terminal point $\bm{x}(\tau) = \bm{x}_{\tau}$ where $\tau = \min(\{t \in \R: g(\bm{x}(t)) \geq 0\})$. The way in which \ac{SAIS} methods use samples of one step to create samples of the next step can be loosely approximated by the following informal procedure. First, randomly choose a sample from the current step, where the selection is done so that samples with higher performance are more likely to be chosen. Next, travel along the associated solution trajectory of the chosen sample for some amount of time. The point reached on the solution trajectory after this time is a sample for the next step. Iterate this process to generate the necessary number of samples. Loosely speaking, from step to step, \ac{SAIS} methods tend to travel in the direction of the solution trajectories of the samples in the current step with the highest performance.

Other approaches seek to improve the exploration of rare-event regions by exploiting additional geometric or dynamical structure. For example, Hamiltonian Monte Carlo methods have been adapted to Subset Simulation \cite{wangHamiltonianMonteCarlo2019}, and more recent energy-based approaches have also been proposed for rare-event probability estimation \cite{friedliEnergybasedModelApproach2025}. Such methods are complementary to the present work, but they rely on structure that is not assumed to be available here. In contrast, \ac{NIS} is designed for black-box reliability problems in which gradient information is not assumed accessible.

Following the above discussion, two performance functions with challenging topology are considered, both found in \cite{breitungGeometryLimitState2019}. Note that both have been multiplied by $-1$, since this paper defines the failure region as inputs which have performance greater than the critical threshold. The first one is the piecewise linear performance function,
\begin{equation}\label{eq:piecewise linear}
\begin{aligned}
    & g_{\text{pwl}}(\bm{x}) = - \min(g_1(x_1),g_2(x_2)), \,\text{where}\\
    & g_1(x_1) =
    \begin{cases}
        4 - x_1 & x_1 > 3.5, \\
        0.85 - 0.1x_1 & x_1 \leq 3.5,
    \end{cases} \\
    & g_2(x_2) =
    \begin{cases}
        0.5 - 0.1x_2 & x_2 > 2, \\
        2.3 - x_2 & x_2 \leq 2.
    \end{cases}
\end{aligned}
\end{equation}

\begin{figure}
    \centering
    \begin{subfigure}[b]{\textwidth}
        \centering
        \includegraphics[trim={0cm 0.3cm 0cm 0cm},clip, scale=0.7]{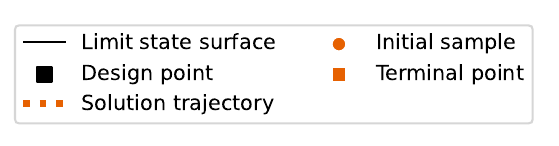}
    \end{subfigure}
    \begin{subfigure}[b]{0.475\textwidth}
        \centering
        \includegraphics[scale=0.55]{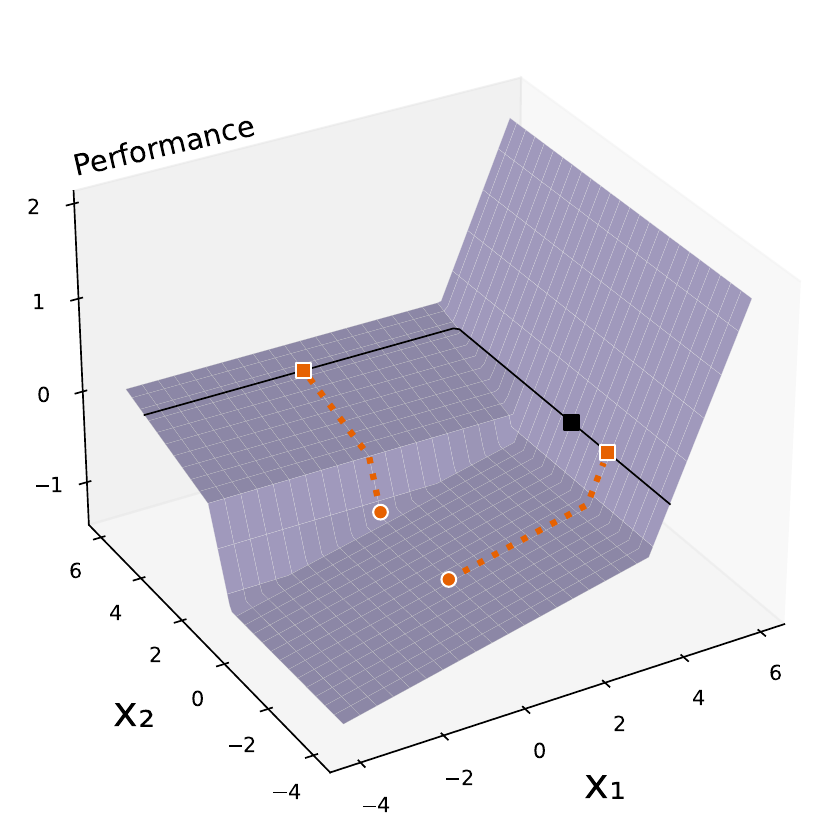}
        \caption{Piecewise linear function.}
        \label{fig:3d_pwl}
    \end{subfigure}
    \hfill
    \begin{subfigure}[b]{0.475\textwidth}
        \centering
        \includegraphics[scale=0.55]{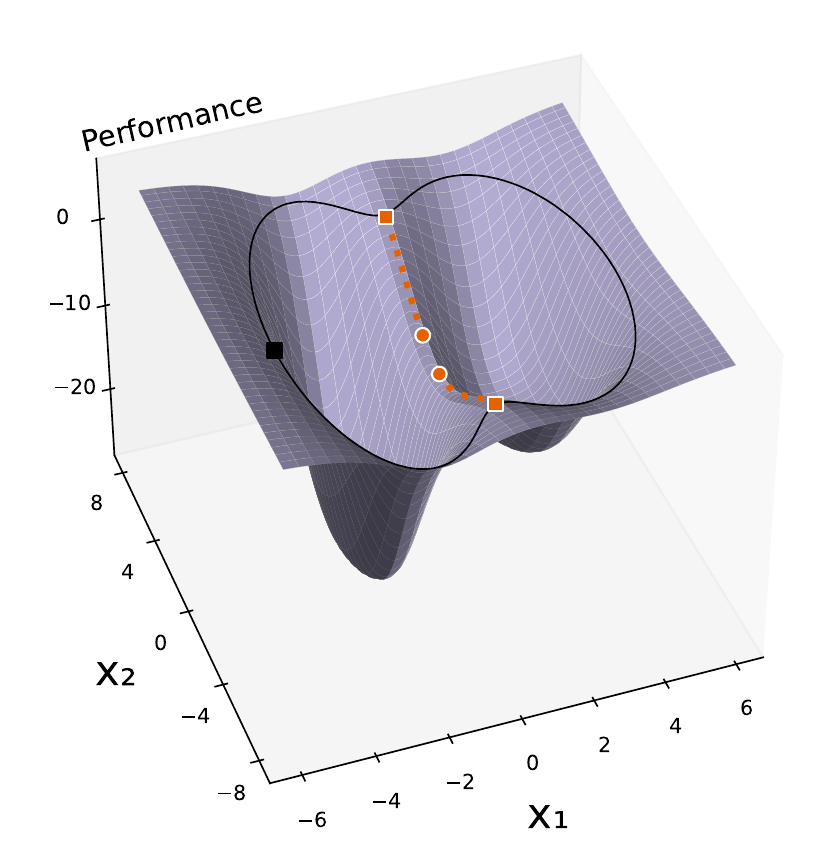}
        \caption{Meatball function.}
        \label{fig:3d_meatball}
    \end{subfigure}
    \hfill
    \caption{Solution trajectories of points sampled from the input distribution.}
    \label{fig:3d}
\end{figure}

\begin{figure}
    \centering
    \begin{subfigure}[b]{\textwidth}
        \centering
        \includegraphics[trim={0cm 0.7cm 0cm 0cm},clip, scale=0.7]{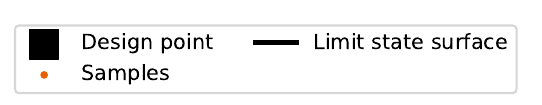}
    \end{subfigure}
    \begin{subfigure}[b]{0.475\textwidth}
        \centering
        \includegraphics[scale=0.45]{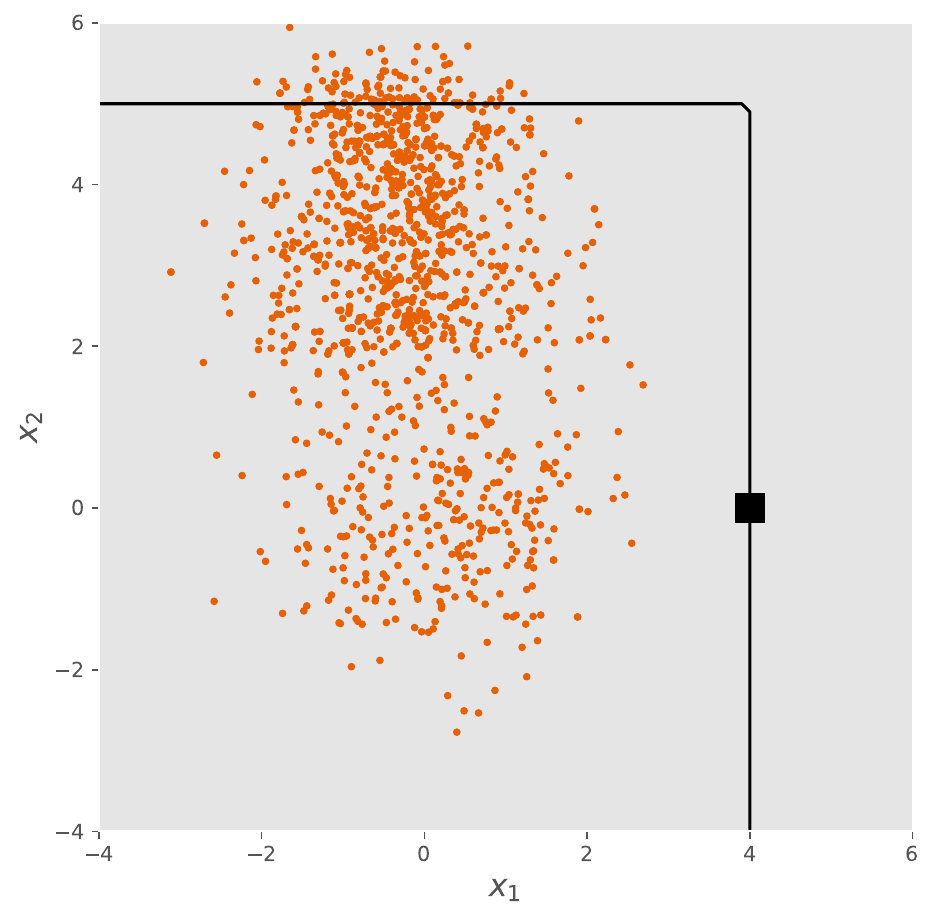}
        \caption{Piecewise linear function.}
        \label{fig:sis_pwl}
    \end{subfigure}
    \hfill
    \begin{subfigure}[b]{0.475\textwidth}
        \centering
        \includegraphics[scale=0.45]{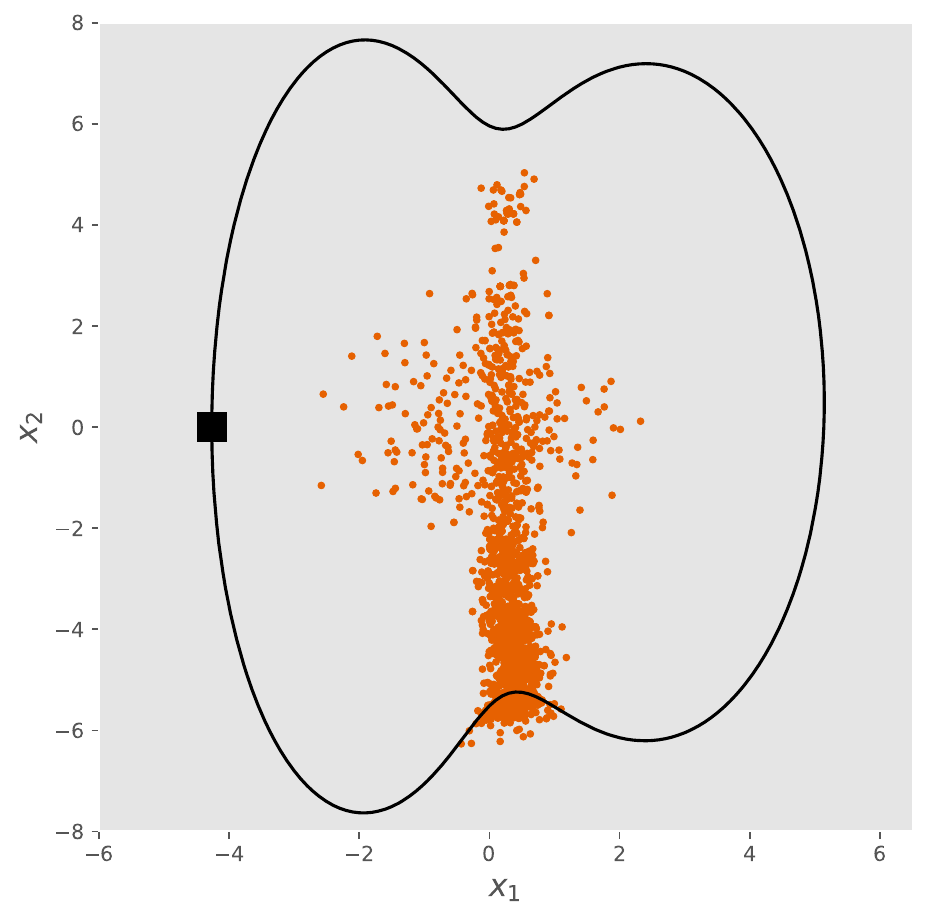}
        \caption{Meatball function.}
        \label{fig:sis_meatball}
    \end{subfigure}
    \hfill
    \caption{Degenerate \ac{SIS} runs on performance functions with challenging topology.}
    \label{fig:sis}
\end{figure}

Figure \ref{fig:3d_pwl} shows a plot of the piecewise linear function, with the reliability problem's limit state surface and design point, along with the solution trajectories of two points that have been sampled from the input distribution. Algorithm runs that do or do not produce samples in the neighbourhood of the design point will be referred to as non-degenerate and degenerate respectively. Note that one of the solution trajectories is degenerate and the other is non-degenerate. Due to the topology of the piecewise linear function, samples from the input distribution with degenerate solution trajectories will tend to have a relatively high performance. According to the above heuristic, this implies that \ac{SAIS} methods acting on the piecewise linear function will struggle to produce samples in the neighbourhood of the design point. Figure \ref{fig:sis_pwl} shows a degenerate \ac{SIS} run on the piecewise linear function.

\begin{figure}
    \centering
    \includegraphics[scale=0.5]{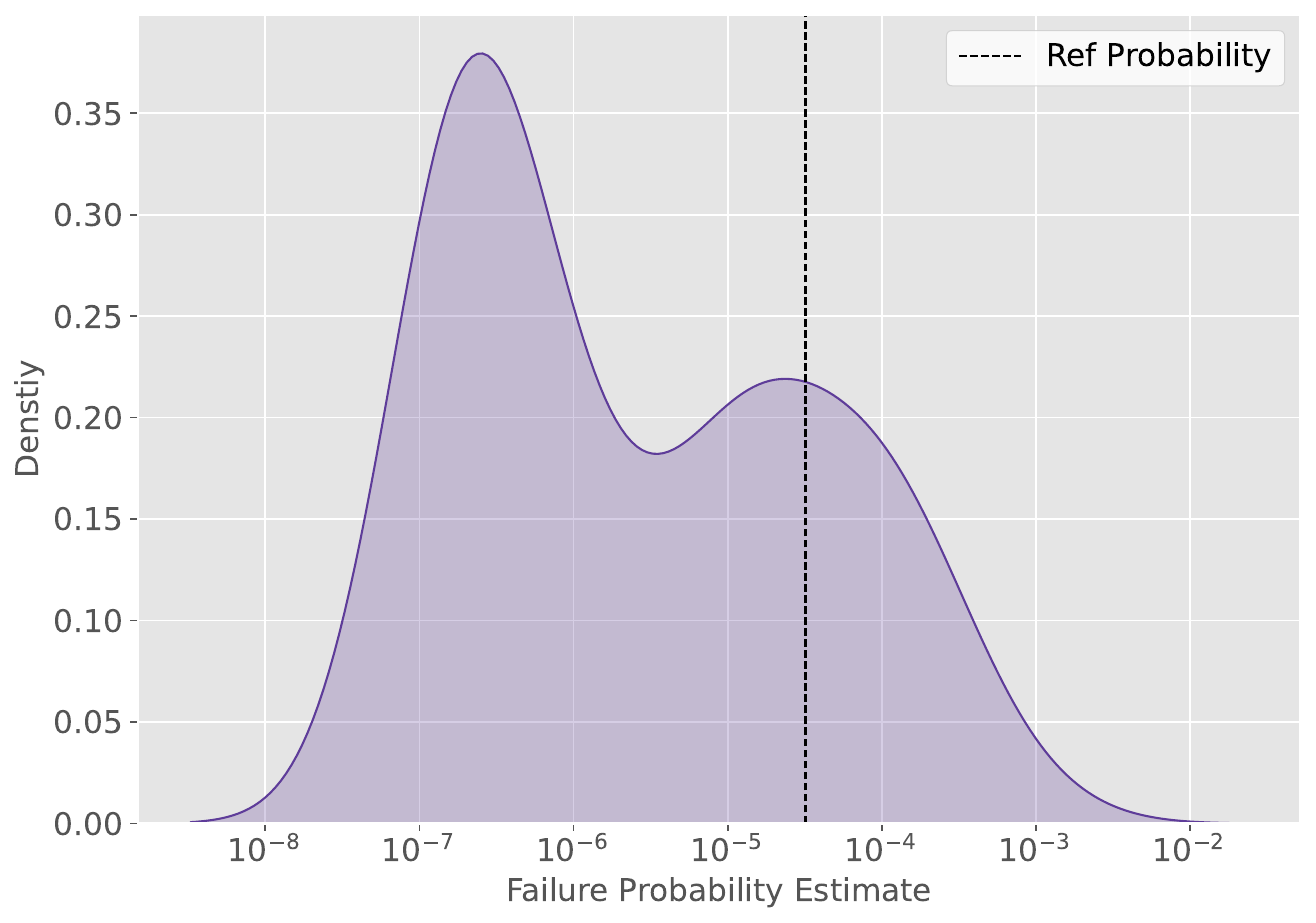}
    \caption{Kernel density estimate of the \ac{SIS} estimator for the piecewise linear function.}
    \label{fig:sis_kde}
\end{figure}

It should be noted that \ac{SAIS} methods are stochastic algorithms. Thus, when acting on challenging performance functions, some percentage of the time they will produce non-degenerate runs, and the rest of the time they will produce degenerate runs. It follows that the probability density of a \ac{SAIS} estimator in these cases is bi-modal, with one mode corresponding to the degenerate case, and the other mode corresponding to the non-degenerate case. Figure \ref{fig:sis_kde} shows a kernel density estimate of 100 \ac{SIS} estimators acting on the piecewise linear function, along with a correct reference probability of failure. Despite the fact that the mean of the estimators may be an acceptable estimate for the probability of failure, the variance of the distribution makes the \ac{SIS} estimator in this situation impractical.

The second performance function considered is the meatball function,
\begin{equation}\label{eq:meatball}
\begin{aligned}
    g_{\text{mb}}(x_1, x_2) =
    -\frac{30}{\left( \frac{4(x_1 + 2)^2}{9} + \frac{x_2^2}{25} \right)^2 + 1} \\
    - \frac{20}{\left( \frac{(x_1 - 2.5)^2}{4} + \frac{(x_2 - 0.5)^2}{25} \right)^2 + 1}
    + 5.
\end{aligned}
\end{equation}
Figure \ref{fig:3d_meatball} shows a plot of this function, with the limit state surface and design point,  along with the solution trajectories of two points that have been sampled from the input distribution. In this case, it is very unlikely that any sample from the input distribution will have a non-degenerate solution trajectory. This is due to the local minima between the origin and the design point. This property distinguishes the meatball function as a more challenging performance function than the piecewise linear function for \ac{SAIS} methods. If the algorithm is not even able to produce one sample with a non-degenerate solution trajectory, it has very little chance of populating the neighbourhood of the design point. Figure \ref{fig:sis_meatball} shows a degenerate \ac{SIS} run using the meatball function.

\ifSubfilesClassLoaded{%
    \newpage
    \bibliography{references}%
}

\end{document}

\section{Niching Initial Sampling (NInitS)}\label{sec:niching initial sampling}

\subsection{Niching}\label{sec:niching}

Informally, let a \ac{LHDR} be a subset of the failure region with the following property: if a Markov chain with the optimal importance density as its stationary distribution is started within a \ac{LHDR}, then it is highly likely to remain in that \ac{LHDR} as it travels through the input space. A \ac{LHDR} should be minimal in the sense that it should not contain a \ac{LHDR} as a subset. In evolutionary multi-modal optimisation, a niche generally refers to a neighbourhood of a local optimum. In the reliability analysis context, a \ac{LHDR} is analogous to a local optimum. Again, informally, let a niche of an \ac{LHDR} be a subset of the failure region such that if a Markov chain with the optimal importance density as its stationary distribution is started within it, then the Markov chain would be highly likely to enter the \ac{LHDR} as it travels through the input space. For a given reliability problem, some niches may contribute significantly to the failure probability, whilst others may contribute only marginally. These will be referred to as \textit{important} and \textit{unimportant} niches respectively. If a reliability method is able to sample from and model all the \acp{LHDR} of the important niches, then it should produce a robust estimator for the probability of failure.

Given a set of samples and some objective function, a niching technique can be used to determine which samples belong to which niche. Many niching techniques rely on the Euclidean metric \cite{liSeekingMultipleSolutions2017}. Unfortunately this metric loses utility in high dimensions \cite{klawonnWhatAreClusters2015a}, hence a lot of research is focussed on low-dimensional objective functions \cite{liBenchmarkFunctionsCEC2013}. These techniques would not be satisfactory, since the goal is to make \ac{NInitS} applicable in high dimensions.

Hill valley tests are a class of niching method that detect valleys in the objective function between two samples \cite{ursemMultinationalEvolutionaryAlgorithms1999}. The idea is that two points belong to two separate niches if they exist on two hills of an objective function, separated by a valley. Most hill valley tests are able to perform well in high dimensions. However, they do require additional objective function evaluations. In this work, the following simple hill valley test is defined for two samples $\bm{x},\bm{y} \in \R^d$,
\begin{equation}\label{eq:hvt}
    \chi(\bm{x},\bm{y}) =
    \begin{cases}
        1 & \text{if } g\left(\frac{\bm{x} + \bm{y}}{2}\right) \geq \min(g(\bm{x}),g(\bm{y})),\\
        0 & \text{otherwise.}
    \end{cases}
\end{equation}
Note that a $1$ is returned when no valley is detected. This midpoint-based test is intentionally parameter-free and is used here as a low-cost heuristic for distinguishing niches without introducing additional user-defined tuning parameters. More elaborate hill valley tests could be considered, but they would generally require additional performance evaluations and are therefore left for future work. This hill valley test will be useful, as it allows \ac{NInitS} to avoid exploring a niche it has already populated. The motivation for the \ac{NInitS} procedure is that it should ideally produce exactly one sample in each important niche. Any more than one sample per important niche would likely be a waste of computational resources.

\subsection{Initial Sampling}\label{sec:initial sampling}

The purpose of an initial sampling procedure in \ac{IS} reliability methods is to produce failure samples that can be used to fit an importance distribution directly or, as is the case for the proposed \ac{NIS}, Markov chain seeds. The most robust initial sampling procedure is to sample from the input distribution and to reject samples not in the failure region \cite{bucherAdaptiveSamplingIterative1988}. Of course, for small failure probabilities this approach is inefficient. Though not explicitly a distinct initial sampler, the first step in \cite{rashkiSESCNewSubset2021} entails sampling from a scaled input distribution. This is only effective if the scaling is able to increase the probability of failure. The optimisation method in \cite{derkiureghianMultipleDesignPoints1998} can efficiently locate multiple niches by adding a bulge to the limit state function. However, it does require gradient information. The technique in \cite{melchersSearchbasedImportanceSampling1990} is gradient-free, though in the case of multiple niches, it requires prior expert knowledge regarding the performance function.

The purpose of \ac{NInitS} is not to generate samples from a single target distribution. Rather, it is a heuristic exploration procedure whose output is a set of representative failure samples intended to seed Markov chains in distinct important niches of the failure region.

More recently, it is common to use a \ac{SIS} algorithm for the initial sampling. For instance, \cite{mehniReliabilityAnalysisCrossentropy2023} uses \ac{SuS} for initial sampling and \cite{tabandehReviewAssessmentImportance2022a} uses a density extrapolation approach combined with \ac{SuS} \cite{jiaDensityExtrapolationApproach2021}. As discussed previously, \ac{SIS} methods can fail to populate all of the important niches. Another approach taken in \cite{tabandehReviewAssessmentImportance2022a} to deal with challenging performance functions is to sample uniformly within some predetermined bounded area. This would not be a viable strategy for a reliability problem with a high-dimensional input space.

The \ac{NInitS} procedure can be thought of as a series of \ac{SuS} runs, where each run only uses one chain per step. These runs will henceforth be referred to as \textit{chain runs}. To begin, one initial seed is sampled from the input distribution. Next, a Markov chain is started from that seed and run until it produces $p^{-1} \in \mathbb{N}$ samples, where $p \in (0,1)$ is a user defined parameter known as the level probability. The Markov chain should have $\eta^1_0$ as its stationary distribution, where the initial intermediate threshold is $b_0 = -\infty$. The performance is then evaluated for all of the samples, and the sample $\bm{x}'$ with the highest performance is identified. The sample $\bm{x}'$ is then chosen as the seed for the next Markov chain, this time with stationary distribution $\eta^1_1$, determined by $b_1 = g(\bm{x}')$. Once more, this new chain produces $p^{-1}$ samples. The chain run continues this way until a stopping condition is triggered. The algorithm then samples an initial seed from the input distribution and the entire process restarts.

Within each chain run, the samples are generated by Markov chains whose stationary distributions are the successive intermediate failure densities. However, the collection of samples produced across all runs is not interpreted as a sample from one common distribution; instead, it is used to identify and represent distinct important niches.

The Markov chain algorithm used in this paper is the Modified Metropolis Algorithm, proposed in the original SuS paper \cite{auEstimationSmallFailure2001}. This algorithm is specifically designed to be efficient in high dimensions. It does this by taking advantage of the structure of the intermediate failure densities, assuming that the input distribution is the standard multivariate normal. Specifically, it takes advantage of the fact that the input dimensions are independent, that is, $f(x)= \prod_{i=1}^d f_i(x_i)$ where the $f_i(x_i)$ are standard univariate normal probability density functions. The Modified Metropolis Algorithm requires a proposal distribution $\xi(\cdot|x)$, dependent on the current state of the chain, $x$. The implementation used for this paper uses the normal distribution, where $\sigma$ is a user-defined proposal scale parameter. The process is summarised in Algorithm \ref{alg:modified metropolis}. Note that even if the input distribution is not the standard multivariate normal, \ac{NInitS} is still applicable, as long as a Markov chain algorithm that can sample efficiently from the resulting intermediate densities can be identified.

\begin{algorithm}
\caption{Modified Metropolis Algorithm}\label{alg:modified metropolis}
\textbf{Input} \\
Sample: $\bm{x} \in \R^d $ \\
\textbf{Parameters} \\
Stationary distribution: $f_s \propto \one_{F}(\bm{x}) \prod_{i=1}^d f_{i}(x_i)$ \\
Proposal distribution: $\xi(\cdot|x): \R^{d} \rightarrow \R$
\begin{algorithmic}[1]
\Procedure{ModifiedMetropolis}{$\bm{x}; f_s$}
    \For{$1\leq i \leq d$}
        \State Sample $ x'_{i} \sim \xi(\cdot|x_i)$
        \State $\theta \gets \min(1, f_i(x'_i)/f_i(x_i))$
        \If{$\textsc{bernoulli}(\theta) = 0$}
            \State $x'_{i} \gets x_i $
        \EndIf
    \EndFor
    \State $\bm{x}' \gets (x'_1,\dots,x'_d)$
    \If{$\one_{F}(\bm{x}') = 0$}
        \State $\bm{x}' \gets \bm{x} $
    \EndIf
    \State \textbf{return} $\bm{x}'$
\EndProcedure
\end{algorithmic}
\end{algorithm}

A Markov chain run can be stopped in any of the following scenarios. Firstly, if the chain contains any samples in the failure region. Secondly, in some cases, a chain run might not be able to produce failure samples in a reasonable amount of steps. This could happen if the chain run becomes stuck in a local maximum of the performance function \cite{valdebenitoRoleDesignPoint2010}. To account for this, a chain run is stopped if $b_m = b_{m-n_\text{con}}$ where $b_m$ is the latest defined intermediate threshold and $n_\text{con}$ is a user defined parameter known as the convergence limit. Thirdly, a chain might make very slow progress towards the failure region, likely implying the region of the failure region it is approaching makes a negligible contribution to the probability of failure. If a chain run consists of more than $n_\text{len}$ chains, then it is stopped, where the length limit $n_\text{len}$ is a user defined parameter.

Let $\bm{X}$ denote a chain run and let $\textsc{ChainStop}(\bm{X})$ denote any combination of the three previously described stopping conditions. That is, $\textsc{ChainStop}(\bm{X})$ returns $\textbf{True}$ if any of the stopping conditions are triggered, and  $\textbf{False}$ otherwise. When a chain run is stopped and it contains samples in the failure region, the failure sample that was last to be generated is added to a set of initial samples, and a set of representatives, denoted $\mathcal{I}$ and $\mathcal{R}$ respectively. Note that $\mathcal{I}$ is the final output of the entire \ac{NInitS} algorithm. If a chain run is stopped and there are no samples in the failure region, then the sample with the highest performance is added to the set of representatives.

The hill valley test can be used to define a hill valley niche for a point $\bm{x} \in \R^d$,
\begin{equation}\label{eq:hvn}
    \chi^{\bm{x}} = \{ \bm{y} \in \R^d: \chi(\bm{x},\bm{y}) =1 \}.
\end{equation}
Together with the set of representatives, the hill valley niche can be used to define the admissible region of the input space,
$\mathcal{A} = \R^d \setminus \bigcup\limits_{\bm{x} \in \mathcal{R}} \chi^{\bm{x}}$
. The admissible region $\mathcal{A}$ represents the subset of the input space that \ac{NInitS} has not yet explored. For the entire run of the algorithm, if a sample is not in $\mathcal{A}$, then it is rejected. If the initial seed sampled from the input distribution is rejected, then a new initial seed is sampled from the input distribution. If a sample from a Markov chain is rejected for not being in $\mathcal{A}$, then the chain returns to its previous state.

Let $\{\sigma_{i}\}_{i=1}^{n_\text{noise}}$ be a user-defined noise sequence of increasing scales,  i.e. $\sigma_{i} < \sigma_{i+1}$. When an initial seed is sampled from the input distribution, it may be rejected for not being in the admissible region. When this occurs, the algorithm will sample a new point from the input distribution. On the $i^\text{th}$ attempt at producing an initial seed, the algorithm adds noise, $\mathcal{N}(\bm{0},I\sigma_i^2)$, where $I$ is the $n$-dimensional identity matrix. After $n_\text{noise}$ attempts, the entire \ac{NInitS} algorithm terminates and returns the initial samples it found. Note that the sequence resets after every successful attempt. The purpose of adding noise is to prevent the algorithm from becoming stuck in a hill valley niche that dominates the high-density region of the input distribution. The algorithm can also terminate if the maximum number of initial samples, a user defined parameter denoted as $\mathcal{I}_{max}$, is ever surpassed. The \ac{NInitS} procedure is summarised in Algorithm \ref{alg:niching initial sampling}. For the numerical examples in this work, if the end of noise sequence is reached, and there are no initial samples detected, the noise sequence is simply restarted.

\captionsetup{type=algorithm}
\captionof{algorithm}{Niching Initial Sampling}\label{alg:niching initial sampling}
\textbf{Input} \\
Performance function: $g\colon\R^{d} \rightarrow \R$ \\
\textbf{Parameters} \\
Level probability: $p \in (0,1]$ \\
Input distribution: $f\colon\R^{d} \rightarrow \R$ \\
Noise sequence: $\sigma_i > 0$ for $1 \leq i \leq n_\text{noise}$ \\
Converge limit: $n_{\text{con}} \in \N $ \\
Length limit: $n_{\text{len}} \in \N $ \\
Maximum initial samples: $\mathcal{I}_{\text{max}} \in \N$ \\
\textbf{Subroutines} \\
\textsc{ModifiedMetropolis}: $\R^d \rightarrow \R^d$ \\
\textsc{ChainStop}$(\bm{X}) \in \{ \textbf{True},\textbf{False}\}$
\begin{algorithmic}[1]
\Procedure{NichingInitialSampling}{$g$}
    \State $n \gets p^{-1}$
    \State $\mathcal{R},\mathcal{I} \gets \emptyset, \emptyset$
    \While{$|\mathcal{I}| \leq \mathcal{I}_{max}$}
        \State $\mathcal{A} \gets \R^d \setminus \bigcup\limits_{\bm{x} \in \mathcal{R}} \chi^{\bm{x}}$
        \State $\textsc{SeedFound} \gets \textbf{False}$
        \For{$ 1 \leq i \leq n_{\text{noise}} $}
            \If{$\textsc{SeedFound} \text{ is } \textbf{False}$}
                 \State $\bm{x}^1_1 \sim f + \mathcal{N}(\bm{0},I\sigma_i^2)$
                 \If{$\one_{\mathcal{A}}(\bm{x}_1^1) = 1$} $\textsc{SeedFound} \gets \textbf{True}$
                \EndIf
            \EndIf
        \EndFor
        \If{$\textsc{SeedFound} \textbf{ is False}$} \textbf{return} $\mathcal{I}$
        \EndIf
        \State $F', \bm{X}, k \gets \R^d,\emptyset,0$
        \While{$\textsc{ChainStop}(\bm{X};n_{\text{con}}, n_{\text{len}}) \text{ is } \textbf{False} $}
            \State $k \gets k +1$
            \State $f_s \gets f(\bm{x})  \one_{F' \cap \mathcal{A}}(\bm{x})$
            \For{$ 2 \leq i \leq n $}
                \State $\bm{x}^{k}_{i} \gets \textsc{ModifiedMetropolis}(\bm{x}^{k}_{i-1};f_s)$
            \EndFor
            \State $\bm{X} \gets ((\bm{x}^j_i)_{i=1}^{n})_{j=1}^k$
            \State $i^* \gets \argmax_i \{ g(\bm{x}^k_i)\}$
            \State $b' \gets g(\bm{x}^k_{i^*}) $
            \State $ F' \gets \{\bm{x} \in \R^{d}: g(\bm{x}) \geq b'\}$
            \State $\bm{x}^{k+1}_1 \gets \bm{x}^{k}_{i^*} $
        \EndWhile
        \If{$b' \geq b$}
            \State $i^* \gets \max \{j:g(\bm{x}^k_i) \geq b  \}$
            \State $\mathcal{I} \gets \mathcal{I} \cup \{\bm{x}^k_{i^*} \}$
            \State $\mathcal{R} \gets \mathcal{R} \cup \{\bm{x}^k_{i^*} \}$
        \Else
            \State $\mathcal{R} \gets \mathcal{R} \cup \{\bm{x}^{k+1}_1  \}$
        \EndIf
    \EndWhile
    \State \textbf{return} $\mathcal{I}$
\EndProcedure
\end{algorithmic}

\begin{figure}
    \centering
    \begin{subfigure}[b]{\textwidth}
        \centering
        \includegraphics[trim={0cm 0.7cm 0cm 0cm},clip, scale=0.7]{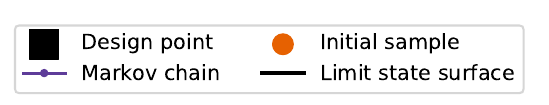}
    \end{subfigure}
    \begin{subfigure}[b]{0.475\textwidth}
        \centering
        \includegraphics[scale=0.45]{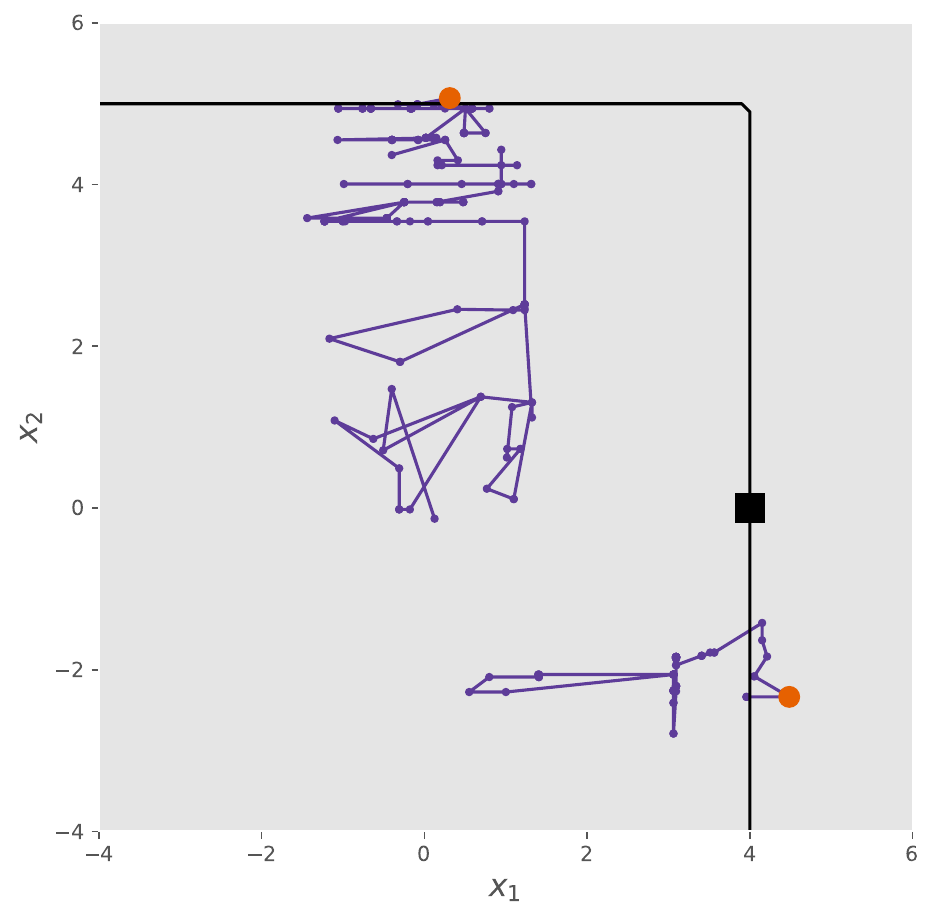}
        \caption{Piecewise linear function.}
        \label{fig:nis_pwl_initial}
    \end{subfigure}
    \hfill
    \begin{subfigure}[b]{0.475\textwidth}
        \centering
        \includegraphics[scale=0.45]{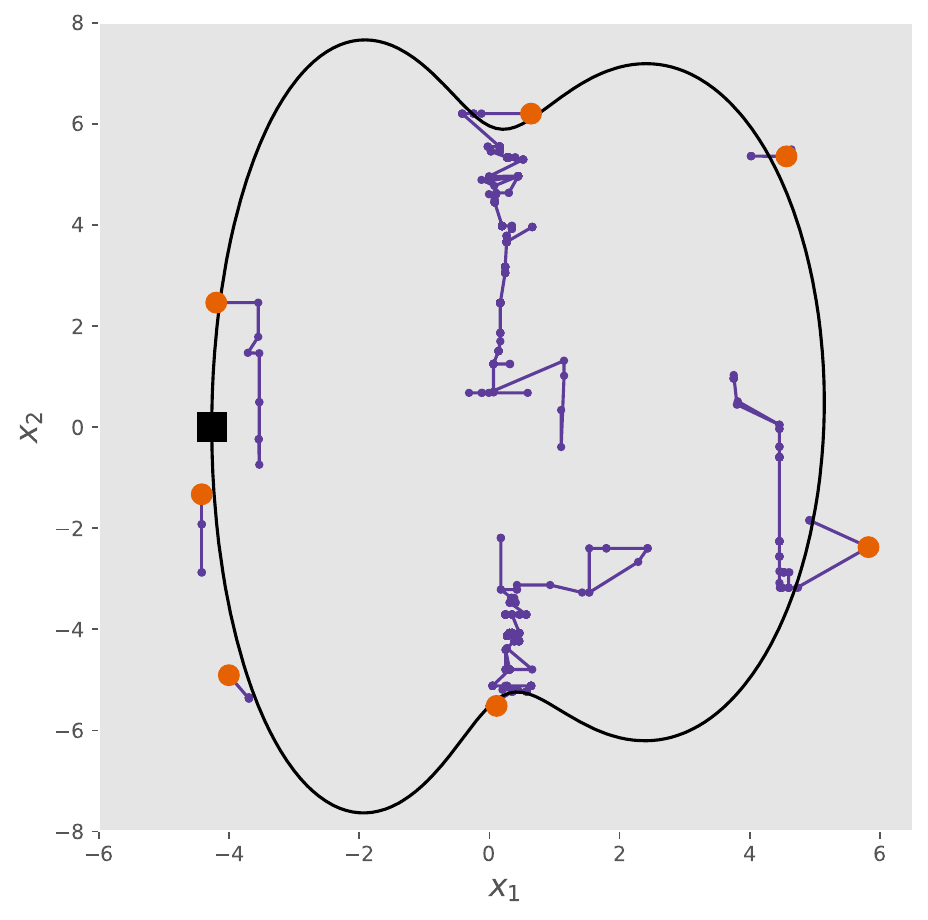}
        \caption{Meatball function.}
        \label{fig:nis_meatball_initial}
    \end{subfigure}
    \hfill
    \caption{Chain runs of niching initial sampling procedure for challenging performance functions.}
    \label{fig:nis_initial}
\end{figure}

Figure \ref{fig:nis_initial} shows the result of running the \ac{NInitS} procedure on the piecewise linear function and the meatball function introduced in Section \ref{sec:importance sampling}. The piecewise linear function has two niches, at the top and the right of Figure \ref{fig:nis_pwl_initial}. The meatball function has four niches, at the top, bottom, left and right of Figure  \ref{fig:nis_meatball_initial}. Since the piecewise linear function is relatively simpler, the niching initial sampler is able to produce exactly one initial sample in each of the two niches. Due to the relatively complex limit state surface of the meatball function, the \ac{NInitS} procedure yields more than one initial sample in some of the niches. In the final mixture importance distribution described in the next section, initial samples in the same niches will correspond to overlapping mixture components. Note also that some of the initial samples are the result of very short chain runs. This is because the initial seed of those chain runs likely had a lot of noise applied to them.

\ifSubfilesClassLoaded{%
    \newpage
    \bibliography{references}%
}

\end{document}

\section{Niching Importance Sampling (NIS)}\label{sec:niching importance sampling}

The \ac{NInitS} procedure introduced in Section \ref{sec:niching} is designed to be integrated into \ac{NIS}, an importance sampling framework. After \ac{NInitS} generates initial samples, the algorithm proceeds iteratively, and each iteration comprises three steps. First, Markov chains are run, starting from each of the initial samples. Next, an importance distribution is fit using the Markov chain samples. Finally, importance samples are generated from the importance distribution and used to estimate the probability of failure. This process is repeated until a stopping condition is triggered and the algorithm terminates. On some of the iterations, depending on a condition that will be described later, the Markov chains are not updated and no new importance distribution is fit. Instead, the importance distribution from the previous iteration generates an additional batch of importance samples that are combined with the previous batch, in order to yield a more accurate estimate for the probability of failure.

\subsection{Updating Markov Chains}\label{sec:updating}

Let $K_{\text{init}}=|\mathcal{I}|$ initial samples be used as seeds for Markov chains with the optimal importance density as the stationary distribution. Modified Metropolis is again used as the Markov chain algorithm for this step. The $K_{\text{init}}$ chains will be denoted as $((\bm{x}^k_i)_{i=1}^{n_k})_{k=1}^{K_{\text{init}}}$, where $n_k$ denotes the length of the $k^{\text{th}}$ chain. On the first iteration, immediately after \ac{NInitS} has finished, $n_k = 1$  for $1 \leq k \leq K_{\text{init}}$, since each chain consists of only an initial sample. Some care must be taken when deciding how many steps each Markov chain should take. A large amount of steps could be needlessly computationally expensive, whereas too few could lead to the importance distribution being a poor approximation of the optimal importance density. There are a few factors to consider.

The first factor is the dimension of the problem. In general, the higher the dimension of the reliability problem, the more samples will be required in order to fit an importance distribution. The second factor is the number of important niches. Since samples from one niche offer very little information about any other niche, a reliability problem with more important niches will require more samples. The estimated number of effective niches, denoted $\hat{K}_{\text{eff}}$, will quantify this idea. Finally, some chains may be exploring regions of higher density than others, and should be assigned more steps. In a problem with only one niche, this would not be necessary, since the Markov chains would automatically distribute the samples correctly across the failure region. However, with multiple niches, the chains experience ergodic problems, so they may sample different areas of the failure region disproportionate to their contribution to the probability of failure. Let the chain weights be denoted as $\alpha_1,\dots,\alpha_{K}$ with $\sum_{k} \alpha_k =1$.

To incorporate all this information into one heuristic, first the total budget is calculated,
\begin{equation}\label{eq:total budget}
    T = M \cdot \hat{K}_{\text{eff}} \cdot \max(d,25),
\end{equation}
where $M$ is a user-defined parameter called the budget multiplier. The dimension component is limited from below by $25$, since for low dimensions this heuristic was empirically found to be too small otherwise. The $k^{\text{th}}$ chain is then updated by $\lfloor \alpha_k T \rfloor$ steps. In the first iteration, there is not enough information to accurately determine the number of effective niches and the chain weights, thus they are set conservatively as $\hat{K}_{\text{eff}} = 1$ and $\alpha_k = 1/K_{\text{init}}$ for $1 \leq k \leq K_{\text{init}}$. The reason the estimated number of effective niches is not set to $K_{\text{init}}$ is that multiple initial samples may be in the same niche, and not all the populated niches may be important. Information gained during the first iteration will be used to set these parameters more accurately for the second iteration. Though not considered in this work, the effective number of samples each chain produces could also be used to help determine the total budget.

\begin{figure}
    \centering
    \begin{subfigure}[b]{\textwidth}
        \centering
        \includegraphics[trim={0cm 0.7cm 0cm 0cm},clip, scale=0.7]{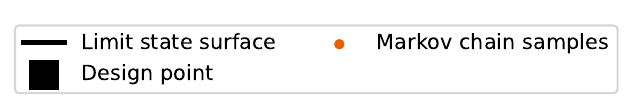}
    \end{subfigure}
    \begin{subfigure}[b]{0.475\textwidth}
        \centering
        \includegraphics[scale=0.45]{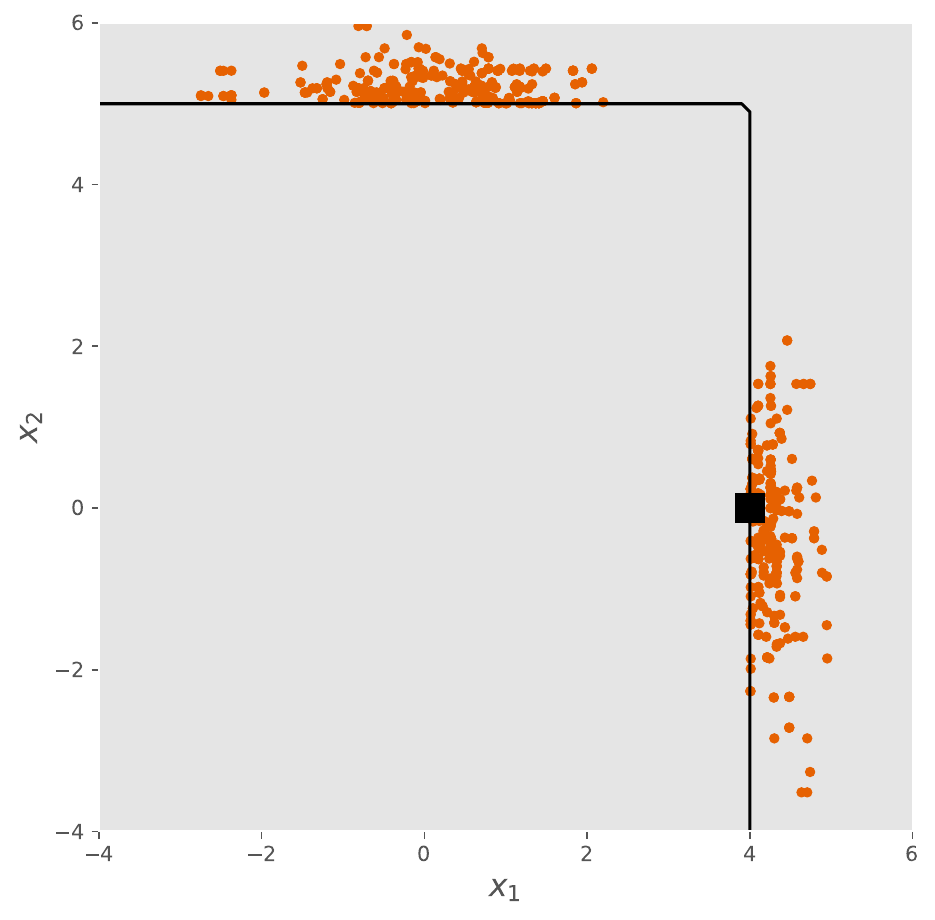}
        \caption{Piecewise linear function.}
        \label{fig:nis_pwl_chain}
    \end{subfigure}
    \hfill
    \begin{subfigure}[b]{0.475\textwidth}
        \centering
        \includegraphics[scale=0.45]{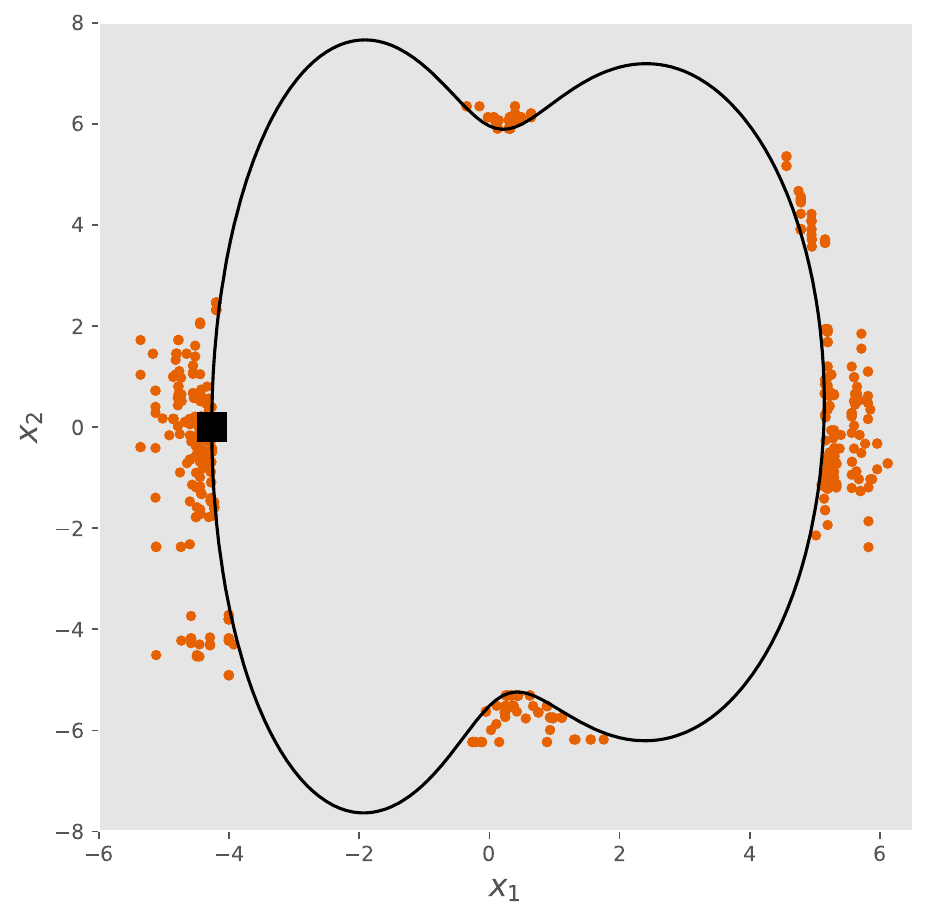}
        \caption{Meatball function.}
        \label{fig:nis_meatball_chain}
    \end{subfigure}
    \hfill
    \caption{Markov chains of the niching importance sampling procedure after one update, applied to challenging performance functions.}
    \label{fig:nis_chain}
\end{figure}

Figure \ref{fig:nis_chain} shows the result of updating the Markov chains for the first time on the meatball and piecewise linear performance functions. Notice that in both cases, the neighbourhood of the design point has been populated. However, due to the ergodicity problems, there are also many samples in unimportant niches in both examples. At this stage, \ac{NIS} is not aware of the relative importance of each niche. Note also that for the meatball function, though seven initial samples were used as seeds, as shown in Figure \ref{fig:nis_meatball_initial}, it is now more clear in Figure \ref{fig:nis_meatball_chain} that there are only four niches as the chains have begun to overlap. This idea will be captured by the estimation of the effective number of components later in the algorithm.

\subsection{The von Mises-Fisher-Nakagami Mixture (vMFNM)}\label{sec:von}

The choice of the parametric family of probability densities to consider is simplified by the assumption that the input distribution is a standard multivariate normal. If the distance of a point to the origin is referred to as its radius, then in high dimensions the probability density of the standard multivariate normal is highly concentrated around points with radius $\sqrt{d}$ \cite{katafygiotisGeometricInsightChallenges2008}. The fact that most of the density exists on a manifold, sometimes called the important ring, with lower dimensionality than the input space can be exploited to make the process of constructing the importance distribution more efficient. To do this, a parametric family of probability densities can be chosen to naturally align with the important ring. The result is that relatively few parameters are required to describe the parametric family, and consequently, less samples are required to fit those parameters.

This paper follows the approach suggested in \cite{papaioannouImprovedCrossEntropybased2019}, of fitting a \ac{vMFNM} model. It should be noted that this approach is a generalisation of the strategy used in \cite{wangCrossentropybasedAdaptiveImportance2016}, where a \ac{vMFM} model is considered. The \ac{vMFM} model assumes that the failure samples have radius $\sqrt{d}$, which becomes inaccurate in low dimensions. Conversely, the \ac{vMFNM} distribution explicitly models the radius of the failure samples. The \ac{vMFNM} model is most naturally described in polar coordinates. Each $\bm{x} \in \R^d$ can be rewritten as $\bm{x} = r\bm{a}$ where $r \in \R$ is a scalar radius and $\bm{a} \in \R^d$ is a unit direction vector. The direction and radius are modelled separately, by the \ac{vMF} distribution and Nakagami distribution respectively. Although the main motivation for this family is its suitability for high-dimensional standard normal spaces, the additional Nakagami radius model also makes it applicable in the lower-dimensional settings considered in the numerical examples.

The \ac{vMF} distribution is given by,
\begin{equation}\label{eq:vmf}
    f_{\text{vMF}}(\bm{a};\bm{\mu} ,\kappa) = C_d(\kappa) \exp(\kappa \bm{\mu}^T \bm{a}),
\end{equation}
where $\bm{\mu}$ is the mean direction, with $\|\bm{\mu}\| = 1$, and $\kappa \geq 0 $ is the concentration parameter. The normalising constant $C_d(\kappa)$ is defined as
\begin{equation}\label{eq:normalising vmf}
    C_d(\kappa) = \frac{\kappa^{d/2 -1}}{(2\pi)^{d/2}I_{d/2-1}(\kappa)},
\end{equation}
where $I_k$ is the modified Bessel function of the first kind with order $k$. The Nakagami distribution is given by
\begin{equation}\label{eq:nakagami}
    f_{\text{N}}(r;m ,\Omega) = \frac{2m^m}{\Gamma(m)\Omega^m}r^{2m-1}\exp\left(-\frac{m}{\Omega}r^2\right),
\end{equation}
where $m \geq 0.5$ is the shape parameter, $\Omega > 0$ is the spread parameter and $\Gamma$ is the gamma function. Together, these distributions define the \ac{vMFN} distribution,
\begin{equation}\label{eq:vMFN}
    f_{\text{vMFN}}(r,\bm{a};m,\Omega,\bm{\mu},\kappa) = f_{\text{N}}(r;m ,\Omega) \cdot f_{\text{vMF}}(\bm{a};\bm{\mu} ,\kappa).
\end{equation}

Finally, in order to be capable of modelling failure regions with multiple important niches, a mixture distribution is defined. This requires a set of $K \times 5$ parameters,
\begin{equation}\label{eq:params}
    \bm{\nu} =  \{ \bm{\nu}_k =(\pi_k,m_k,\Omega_k,\bm{\mu}_k,\kappa_k): 1 \leq k \leq K\},
\end{equation}
where the $\pi_k$ are the component weights with $\sum_k \pi_k =1$ and $K$ is the number of components. Let a \ac{vMFN} component be defined as
\begin{equation}\label{eq:vMFNc}
    f_{\text{vMFNc}}(r,\bm{a};\bm{\nu}_k) = \pi_k \cdot f_{\text{vMFN}}(r,\bm{a};m_k,\Omega_k,\bm{\mu}_k,\kappa_k).
\end{equation}
The \ac{vMFNM} distribution is given as
\begin{equation}\label{eq:vmfnm}
    f_{\text{vMFNM}}(r,\bm{a};\bm{\nu}) = \sum_{k=1}^{K}f_{\text{vMFNc}}(r,\bm{a};\bm{\nu}_k).
\end{equation}
The \ac{vMFNM} family is not assumed to represent the optimal importance density exactly; rather, it is used as a flexible parametric approximation from which importance samples can be generated efficiently. Any mismatch between the fitted mixture and the optimal importance density therefore affects the efficiency of the resulting estimator, rather than replacing the weighted importance sampling estimate itself.

\subsection{Expectation-Maximisation (EM)}\label{sec:em}

The samples from the Markov chains can be used to fit the mixture distribution. In this section, all the samples from all the Markov chains will be considered as one set of collective chain samples, written in polar coordinates as
\begin{equation}\label{eq:relabel}
    \bigcup_{k=1}^{K_{\text{init}}}\bigcup_{i =1}^{n_k} \bm{x}_i^k =  (r_i\bm{a}_i)_{i=1}^{N}, \text{ where } N = \sum_{k=1}^{K_{\text{init}}} n_k.
\end{equation}
The parameters can be fit to the samples by minimising the relative cross entropy between the importance distribution and the optimal importance density. Following the cross entropy approach in Section \ref{sec:cross entropy}, an approximate objective function can be derived,
\begin{equation}\label{eq:mle}
    L(\bm{\nu}) =  \frac{1}{N} \sum_{i=1}^N \ln(f_{\text{vMFNM}}(r_i,\bm{a}_i;\bm{\nu})).
\end{equation}
Of course, this optimisation problem is equivalent to maximum likelihood estimation for a \ac{vMFNM} distribution. An \ac{EM} algorithm, which is an iterative algorithm where each iteration comprises an expectation step (E-step) and maximisation step (M-step), is typically used to optimise this type of objective function.

This section follows closely the \ac{EM} approach detailed in \cite{papaioannouImprovedCrossEntropybased2019}. However, since the Markov chains sample directly from the optimal importance density, no importance weights appear in the approximate cross entropy objective function. This means that the task can be considered a regular maximum likelihood estimation problem, rather than a weighted maximum likelihood estimation problem. This is significant, since as has been noted previously \cite{wangCrossentropybasedAdaptiveImportance2016, mehniReliabilityAnalysisCrossentropy2023}, the importance weights can become degenerate in high dimensions. That is, the importance weights can become very large for very few samples, and so only very few points contribute to the estimation of the parameters.

On the $l^{\text{th}}$ iteration of the \ac{EM} algorithm, the estimated parameters are denoted as,
\begin{equation}\label{eq:l params}
    \hat{\bm{\nu}}^{(l)} =  \{ \hat{\bm{\nu}}^{(l)}_k =(\hat{\pi}^{(l)}_k,\hat{m}^{(l)}_k,\hat{\Omega}^{(l)}_k,\hat{\bm{\mu}}^{(l)}_k,\hat{\kappa}^{(l)}_k): 1 \leq k \leq K\}.
\end{equation}
In the $l^{\text{th}}$ E-step, the objective function is evaluated as $L(\bm{\nu}^{(l)})$. Note that the objective function is an approximation of an expectation. Next, the latent component label of each sample is estimated. This assignment can be done with the posterior probabilities,
\begin{equation}\label{eq:posterior}
    \gamma_k(r,\bm{a}; \bm{\nu}) = \frac{f_{\text{vMFNc}}(r,\bm{a};\bm{\nu}_k)}
    {f_{\text{vMFNM}}(r,\bm{a};\bm{\nu})}.
\end{equation}
Let $\gamma^{(l)}_{i,k} = \gamma_k(r_i,\bm{a}_i; \hat{\bm{\nu}}^{(l)})$ for $1 \leq k \leq K$ and $1 \leq i \leq N$. On iteration $l+1$, the M-step attempts to maximise the objective function, given the posterior probabilities from iteration $l$. Using auxiliary variables,
\begin{align}\label{eq:aux}
    \bar{R}_k &= \min\left( \frac{\|\sum_{i=1}^{N} \gamma^{(l)}_{i,k}\cdot\bm{a}_{i}\|}{\sum_{i=1}^{N} \gamma^{(l)}_{i,k}},0.95\right), \\
    \mu_{j,k} &= \frac{\sum_{i=1}^{N} \gamma^{(l)}_{i,k}\cdot r_{i}^j}{\sum_{i=1}^{N} \gamma^{(l)}_{i,k}},
\end{align}
the following update rules were derived in \cite{papaioannouImprovedCrossEntropybased2019}:
\begin{align}
    \hat{\pi}_k^{(l+1)} &= \sum_{i=1}^{N}\gamma^{(l)}_{i,k},\\
    \hat{\bm{\mu}}_k^{(l+1)} &= \frac{\sum_{i=1}^{N} \gamma^{(l)}_{i,k}\cdot\bm{a}_{i}}{\|\sum_{i=1}^{N} \gamma^{(l)}_{i,k}\cdot\bm{a}_{i}\|},\\
    \hat{\kappa}_k^{(l+1)} &= \frac{\bar{R}_k(d-\bar{R}_k^2)}{1-\bar{R}_k^2}, \\
    \hat{\Omega}_k^{(l+1)} &= \mu_{2,k}, \\
    \hat{m}_k^{(l+1)} &= \frac{\mu^2_{2,k}}{\mu_{4,k}-\mu_{2,k}^2}, \label{eq:m}
\end{align}
for $1 \leq k \leq K$. The Lagrangian multiplier method can be used to analytically derive the rules for $\pi$, $\mu$ and $\Omega$. This is not possible for $m$ and $\kappa$, and thus the updating rules are approximated using estimators that have been shown to give empirically good results \cite{abdiPerformanceComparisonThree2000,sraShortNoteParameter2012}. The upper limit is placed on $\bar{R}$ for numerical stability. At each step, if $|L(\hat{\bm{\nu}}^{(l+1)}) - L(\hat{\bm{\nu}}^{(l)}) | < 10^{-5} \cdot |L(\hat{\bm{\nu}}^{(l+1)})|$, then the algorithm is stopped, and the final parameter estimator is relabelled as
\begin{equation}\label{eq:final params}
    \hat{\bm{\nu}}^{(l+1)} = \hat{\bm{\nu}} =  \{ \hat{\bm{\nu}}_k =(\hat{\pi}_k,\hat{m}_k,\hat{\Omega}_k,\hat{\bm{\mu}}_k,\hat{\kappa}_k): 1 \leq k \leq K\}.
\end{equation}
For the sake of use in the next section, the final posterior probabilities are also relabelled $\gamma_{i,k} = \gamma_{i,k}^{(l+1)}$.

Note that since a set of parameters is required to compute the posterior probabilities, and the posterior probabilities are required to estimate a set of parameters, some initial set of posterior probabilities must be decided upon starting the \ac{EM} algorithm. Determining the initial posterior probabilities is not typically straightforward and the performance of the \ac{EM} algorithm can be sensitive to this choice. Defining the initial posterior probabilities requires the number of components, $K$, to also be decided upon. If $K$ is too small, then some important niches may be missed. However, if $K$ is too large, there may not be enough samples to fit all the parameters accurately. Even if $K$ can be chosen appropriately, a poor initial assignment of the samples to components could negatively affect the performance of the \ac{EM} algorithm.

In some cases, prior information regarding the reliability problem, such as the number of components, can be used to aid the process. However, in this paper, it is assumed no such information is available. Clustering methods like DBSCAN \cite{esterDensitybasedAlgorithmDiscovering1996} or k-means \cite{macqueenMethodsClassificationAnalysis1967a} could be used to decide the initial posterior probabilities. However, these algorithms have their own user-defined parameters, and will often perform poorly in high dimensions because of their reliance on the Euclidean metric \cite{klawonnWhatAreClusters2015a}. Because \ac{NIS} uses \ac{NInitS}, this task is considerably simplified. Because of \ac{NInitS}, the number of chains, $K_{\text{init}}$, is a reasonable upper bound for the optimal number of components, and so $K$ is set to $K_{\text{init}}$. Additionally, it is natural to initially assign samples from the same chain to the same corresponding component. Formally,
\begin{equation}\label{eq:initial posterior}
    \gamma_{i,k}^{(0)} =
    \begin{cases}
        1 & \text{if } r_i \bm{a}_i \in (\bm{x}_j^k)_{j=1}^{n_k},\\
        0 & \text{otherwise},
    \end{cases}
\end{equation}
for $1 \leq k \leq K$ and $1 \leq i \leq N$.

\subsection{Component Weight Correction}\label{sec:weight}

Because of the ergodicity problems that multiple niches can cause for Markov chains, the component weight estimate from the previous section is likely to be inaccurate. In this section, one last adjustment is made to the component weight estimate, whilst the other estimators are left as they are. One approach could be to use the chain runs from the \ac{NInitS} procedure. The idea is that each chain run is actually a run of \ac{SuS}, hence the estimator $\hat{P}_{\text{SIS}}$ in Equation \ref{eq:sis estimate} could be used to estimate the size of each niche. Unfortunately, due to the fact that only a single chain is used at each step and that the initial seed has noise applied to it, this estimator will also yield an inaccurate result.

Instead, the cross entropy optimisation problem defined in Equation \ref{eq:kl opt_2} is considered with respect to the optimal importance density and with an additional importance weight term,
\begin{equation}\label{eq:cross}
    \bm{\nu}' = \argmax_{\bm{\nu}} \E_{q^*} \left[ \ln(h(\bm{x};\bm{\nu})) \frac{f(\bm{x})}{q^*(\bm{x})}\right].
\end{equation}
The parameter set fit in the previous section can be used to approximate the optimal importance density, $q^*(\cdot) \approx f_{\text{vMFNM}}(\,\cdot\,;\hat{\bm{\nu}})$. This approximation, together with the collective chain samples, may be used to approximate the expectation,
\begin{equation}\label{eq:approx proposal cross}
    \hat{\bm{\nu}}' = \argmax_{\bm{\nu}} \frac{1}{N} \sum_{i=1}^N \ln(f_{\text{vMFNM}}(r_i,\bm{a}_i;\bm{\nu})) w_i,
\end{equation}
with $w_i = f(r_i,\bm{a}_i)/f_{\text{vMFNM}}(r_i,\bm{a}_i;\hat{\bm{\nu}})$ where the input distribution is defined in polar coordinates. The Lagrangian multiplier method may then be used again to derive a weighted updating rule for the component weights,
\begin{equation}\label{eq:weight update}
    \hat{\pi}'_k = \frac{\sum_{i=1}^N w_i \cdot \gamma_{i,k}  }{\sum_{i=1}^N w_i},
\end{equation}
for $1 \leq k \leq K$. Let $\hat{\bm{\nu}}^{\text{IS}}$ denote the parameter set that is the result of replacing $\hat{\pi}_k$ with $\hat{\pi}'_k$ in $\hat{\bm{\nu}}$ for $1 \leq k \leq K$. For use in the next iteration, the chain weights are also updated,
\begin{equation}\label{eq:chain weight update}
    \alpha_k = \frac{\sum_{i=1}^N w_i \cdot \gamma_{i,k}^{(0)} }{\sum_{i=1}^N w_i},
\end{equation}
for $1 \leq k \leq K$.

\subsection{Importance Sampling Estimate}\label{sec:importance sampling estimate}

Finally, the probability of failure is estimated with an \ac{IS} estimator. Importance samples are generated from the importance distribution defined by the parameter set learned in the last section,
\begin{equation}\label{eq:importance samples}
    \{r_i^{\text{IS}}a_i^{\text{IS}}\}_{i=1}^{N_{\text{IS}}} \sim f_{\text{vMFNM}}(\,\cdot\,;\hat{\bm{\nu}}^{\text{IS}}),
\end{equation}
where $N_{\text{IS}}$ is the number of importance samples, a user-defined parameter. The estimator is obtained by substituting the \ac{vMFNM} distribution as the importance distribution into Equation \ref{eq:importance sampling},
\begin{equation}\label{eq:sub importance sampling}
    \hat{P}_{\text{IS}} = \frac{1}{N_{\text{IS}}} \sum_{i=1}^{N_{\text{IS}}} W_i, \text{ with } W_i = \frac{\one_{F}(r_i^{\text{IS}},\bm{a}_i^{\text{IS}}) \cdot f(r_i^{\text{IS}},\bm{a}_i^{\text{IS}})}{f_{\text{vMFNM}}(r_i^{\text{IS}},\bm{a}_i^{\text{IS}};\hat{\bm{\nu}}^{\text{IS}})},
\end{equation}
where the indicator function has been defined in polar coordinates.

\begin{figure}
    \centering
    \begin{subfigure}[b]{\textwidth}
        \centering
        \includegraphics[trim={0cm 0.7cm 0cm 0cm},clip, scale=0.7]{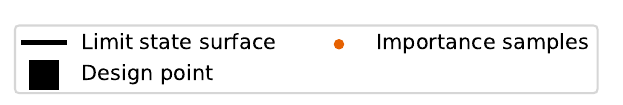}
    \end{subfigure}
    \begin{subfigure}[b]{0.475\textwidth}
        \centering
        \includegraphics[scale=0.5]{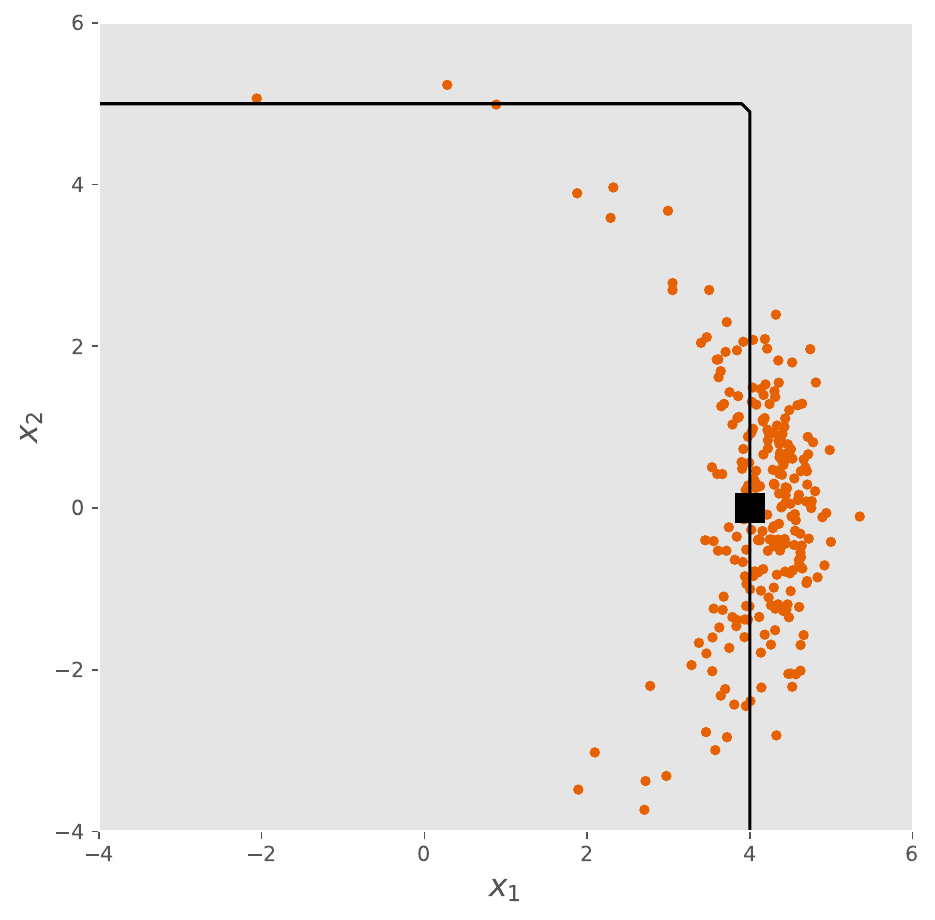}
        \caption{Piecewise linear function.}
        \label{fig:nis_pwl_import}
    \end{subfigure}
    \hfill
    \begin{subfigure}[b]{0.475\textwidth}
        \centering
        \includegraphics[scale=0.5]{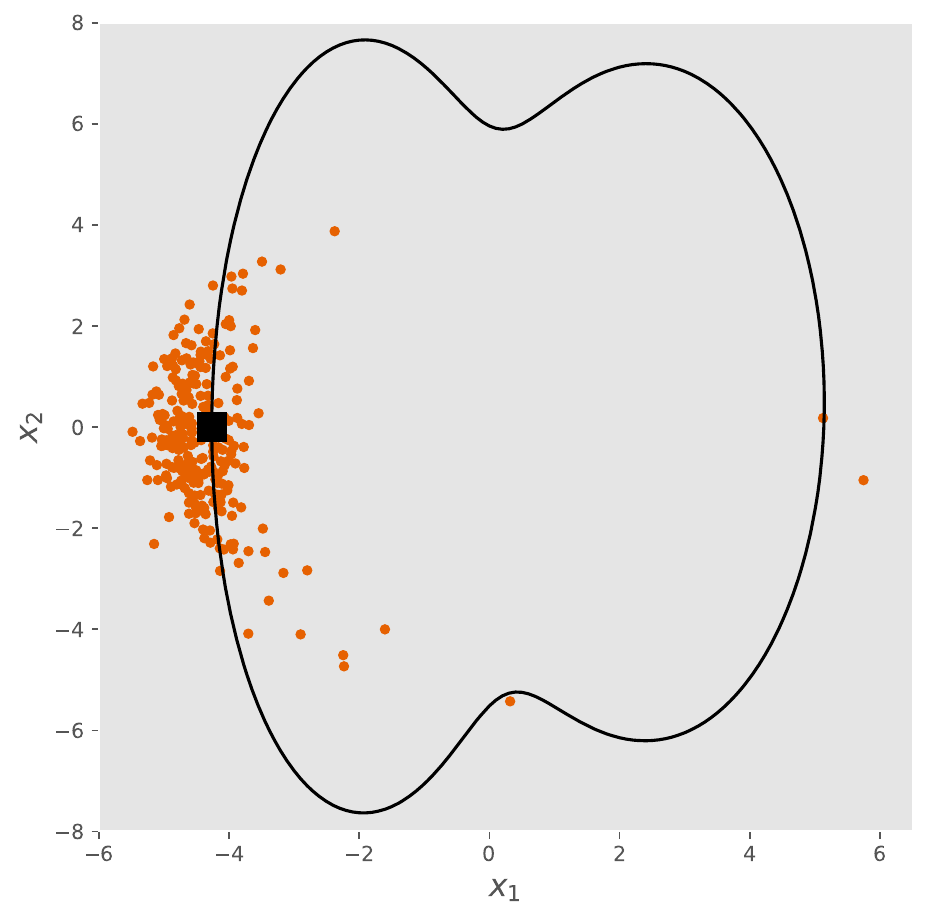}
        \caption{Meatball function.}
        \label{fig:nis_meatball_import}
    \end{subfigure}
    \hfill
    \caption{Importance samples generated by the importance distribution fit using niching importance sampling for two challenging performance functions.}
    \label{fig:nis_import}
\end{figure}

Figure \ref{fig:nis_import} shows the importance samples \ac{NIS} produced when it was run on the piecewise linear function and the meatball function. Note how the vast majority of the samples are in the important niches in both examples, that is, the neighbourhoods of the design points. This is in contrast to the collective Markov chain samples of the first iteration, which are roughly uniformly spread amongst the niches. In later iterations, the chain weights encourage more Markov chain samples to be generated in the important niches.

At this point, the effective number of niches is estimated for use in the next iteration. The mutual information between two random variables is a measure of how much information is gained about one random variable by observing the other. In particular, the mutual information of a \ac{vMFNM} distribution may be defined as
\begin{equation}\label{eq:mutual information}
    \operatorname{MI}\left(\bm{\nu}\right) = \iint f_{\text{vMFNM}}(r,\bm{a};\bm{\nu}) \sum_{k=1}^K \gamma_k(r,\bm{a}; \bm{\nu}) \cdot \ln\left(\frac{\gamma_k(r,\bm{a}; \bm{\nu})}{\pi_k} \right) dr d\bm{a}.
\end{equation}
In this case, the two random variables considered are the latent component labels and the samples from the \ac{vMFNM} distribution. The more uniform the component weights, and the less the components overlap, the higher the mutual information will be. The exponential of the mutual information is bounded between $1$ and $K$, and is often used to represent an effective number of components. In this work, the exponential of the mutual information will be referred to as the effective number of niches, which can be estimated with an \ac{MC} estimator,
\begin{align}
    K_{\text{eff}} &= \exp\left[\operatorname{MI}(\hat{\bm{\nu}}^{\text{IS}})\right] \label{eq:eff_niches}\\
    \hat{K}_{\text{eff}} & = \exp\left[
    \frac{1}{N_{\text{IS}}} \sum_{i=1}^{N_{\text{IS}}} \sum_{k=1}^K \gamma^{\text{IS}}_{i,k} \cdot \ln\left(\frac{\gamma^{\text{IS}}_{i,k}}{\hat{\pi}'_k} \right)
    \right],\label{eq:est_eff_niches}
\end{align}
where $\gamma^{\text{IS}}_{i,k} = \gamma_k(r_i^{\text{IS}}a_i^{\text{IS}}; \bm{\nu}^{\text{IS}})$ for $1 \leq k \leq K$ and $1 \leq i \leq N_{\text{IS}}$.

The control flow of the \ac{NIS} algorithm is determined by two \ac{CoV} estimators,
\begin{align}
    \hat{\delta}_{W} &= \sqrt{\frac{1}{N_{\text{IS}}\cdot \hat{P}^2_{\text{IS}}}\sum_{i=1}^{N_{\text{IS}}} \left(W_i - \hat{P}_{\text{IS}}\right)^2}, \label{eq:weight cov} \\
    \hat{\delta}_{\text{IS}} &= \frac{\hat{\delta}_{W}}{\sqrt{N_{\text{IS}}}},\label{eq:est cov}
\end{align}
that is, the \ac{CoV} estimate of the importance weights and the \ac{CoV} estimate of the \ac{IS} estimator respectively. At the start of each iteration, it is checked whether $\hat{\delta}_{W} > \delta_{w}^\text{target}$, where $\delta_{w}^\text{target}$ is a user-defined parameter. If the condition is true, then the chains and importance distribution are updated. If the condition is false, the chains are not updated and the importance distribution remains the same as in the previous iteration. In this case, the $N_{\text{IS}}$ generated importance samples are combined with the importance samples from the previous iteration. That is, if the algorithm has iterated $l$ times without the importance distribution changing, there will be $l \cdot N_{\text{IS}}$ importance samples. Formally, all references to $N_{\text{IS}}$ in the equations of this section should be replaced with $l \cdot N_{\text{IS}}$. Note that $\hat{\delta}_{W}$ is updated on every iteration, and so at any point the current importance distribution and importance samples can be discarded. The algorithm continues while $\hat{\delta}_{\text{IS}}  > \delta_{\text{IS}}^\text{target}$, where $\delta_{\text{IS}}^\text{target}$ is a user defined parameter. The resulting \ac{NIS} algorithm is summarised in Algorithm \ref{alg:Niching Importance Sampling}.

\begin{algorithm}
\caption{Niching Importance Sampling}\label{alg:Niching Importance Sampling}
\textbf{Input} \\
Performance function: $g\colon\R^{d} \rightarrow \R$ \\
\textbf{Parameters} \\
Optimal importance density: $q^*\colon\R^{d} \rightarrow \R$ \\
Budget multiplier: $M \in \R$ \\
Estimator CoV target: $\delta_{\text{IS}}^\text{target} > 0$ \\
Importance weights CoV target: $\delta_{w}^\text{target} > 0$  \\
Importance sample size: $N_{\text{IS}} \in \Z$ \\
\textbf{Subroutines} \\
\textsc{NichingInitialSampling}$(g)$ \\
\textsc{ModifiedMetropolis}: $\R^d \rightarrow \R^d$
\begin{algorithmic}[1]
\Procedure{NichingImportanceSampling}{g}
    \State $\mathcal{I} \gets \textsc{NichingInitialSampling}(g)$
    \State $K,\,K_{\text{eff}}, \, \hat{\delta}_w, \,
    \hat{\delta}_{\text{IS}}  \gets  |\mathcal{I} |,\, 1,\, \infty,\, \infty$
    \State $\alpha_k = 1/K$ for $1 \leq k \leq K$
    \While{$\hat{\delta}_{\text{IS}}  > \delta_{\text{IS}}^\text{target}$}
    \If{$\hat{\delta}_{w}  > \delta_{w}^\text{target}$}
        \State $T \gets M \cdot K_{\text{eff}} \cdot \max(d,25)$
        \For{$1 \leq k \leq K$}
            \State $n' \gets \lfloor \alpha_k T \rfloor$
            \For{$ 1 \leq i \leq n' $}
                \State $\bm{x}^{k}_{n_{k}+i} \gets \textsc{ModifiedMetropolis}(\bm{x}^{k}_{n_{k}+i-1};q^*)$
            \EndFor
            \State $n_k \gets n_k + n'$
        \EndFor
        \State Compute $\hat{\bm{\nu}}$ with \ac{EM} algorithm using updating Equations \ref{eq:posterior}--\ref{eq:m}
        \State Compute $\hat{\bm{\nu}}^{\text{IS}}$ with Equation \ref{eq:weight update}
        \State Compute $\alpha_k$ for $1\leq k \leq K$ with Equation \ref{eq:chain weight update}
    \EndIf
    \State Sample $N_{\text{IS}}$ times from $f_{\text{vMFNM}}(\,\cdot\,;\hat{\bm{\nu}}^{\text{IS}})$
    \State Compute $\hat{P}_{\text{IS}}$  with Equation \ref{eq:sub importance sampling}
    \State Compute $ \hat{K}_{\text{eff}} $ with Equation \ref{eq:est_eff_niches}
    \State Compute $ \hat{\delta}_w, \hat{\delta}_{\text{IS}} $  with Equation \ref{eq:weight cov} and Equation \ref{eq:est cov}
    \EndWhile
    \State \textbf{return} $\hat{P}_{\text{IS}}$
\EndProcedure
\end{algorithmic}
\end{algorithm}

\ifSubfilesClassLoaded{%
    \newpage
    \bibliography{references}%
}

\end{document}

\section{Numerical Examples}\label{sec:numerical examples}

In this section, \ac{NIS} is compared with \ac{SIS} and \ac{iCE} in a series of numerical experiments. Each algorithm is run $100$ times on each performance function considered. The mean estimate for the probability of failure, the \ac{CoV} of the estimates, and the mean number of performance function evaluations are all reported. Typically, the quality of a reliability method is determined by the number of performance function evaluations required to yield a probability of failure estimator with a given \ac{CoV}. However, it is possible that when a reliability method is applied to certain challenging reliability problems, the resulting estimator is completely degenerate. By \textit{degenerate}, it is meant that either the mean of the estimator deviates significantly from the target reference value, or that the \ac{CoV} is prohibitively large for any reasonable number of performance function evaluations, or both. With the exception of the last example, the performance functions chosen for this section exhibit the type of challenging topology that often causes \ac{SAIS} methods to produce degenerate estimators. The examples are chosen to serve two complementary purposes. The stylised benchmark functions isolate geometric features that are known to be challenging for sequential adaptive methods, while the later application-oriented examples illustrate the behaviour of \ac{NIS} on more practical reliability problems. The results indicate that \ac{NIS} is a robust reliability method that consistently avoids degenerate behaviour on several challenging examples, while remaining effective on problems that do not exhibit the same geometric difficulties.

\ac{NIS} has many tuning parameters that need to be set by the user. In order to demonstrate robustness and to avoid over-fitting the algorithm to each individual reliability problem, these parameters are set to the same constant defaults for all the experiments. The defaults are as follows: the proposal scale, $\sigma=0.8$; the level probability, $p=0.1$; the maximum initial samples, $\mathcal{I}_{\text{max}} = 10$; the chain run convergence limit, $n_{\text{con}} = 20$; the chain run length limit $n_{\text{len}} = 100$; the noise sequence, $(0.00,0.04,\dots,3.96,4.00)$; the budget multiplier, $M=30$; the importance weights \ac{CoV} target, $\delta_{w}^{\text{target}}=5$; the estimator \ac{CoV} target, $\delta_{\text{IS}}^{\text{target}}=0.1$; and the importance sample size, $N_{\text{IS}} = 250$.

Most of the parameters of \ac{SIS} and \ac{iCE} were fixed for the numerical experiments as well. In this paper, neither algorithm's parameters have been discussed in detail, and some parameters have not been discussed at all. For more explicit details, the reader is referred to \cite{papaioannouSequentialImportanceSampling2016,papaioannouImprovedCrossEntropybased2019}. For \ac{SIS}, the adaptive conditional sampler was used along with a level sample size of $2000$, a level probability of $0.1$, a burn-in period of $0$ and a target \ac{CoV} of 1.5. For \ac{iCE}, the vMFNM distribution was used with level size $2000$, maximum iterations set at $30$ and a target \ac{CoV} of 1.5. The number of initial mixture components was varied, and the results of the parameter value that gave the optimal performance are presented.

Given a performance function $g$ with input dimension $d$, it is possible to construct a performance function with identical probability of failure and a higher input dimension $d^*$. Provided $d^*/d =s$ is an integer, the higher dimensional performance function is defined as
\begin{equation}\label{eq:high performance}
    g^*(\bm{x}) = g(z_1,\dots,z_d), \text{ where } z_i = \frac{1}{\sqrt{s}} \sum^{i \cdot s}_{j=(i-1)s +1} \bm{x}_j, \quad \text{ for } 1 \leq i \leq d.
\end{equation}
This construction is used in the numerical examples to embed a low-dimensional benchmark problem into a higher-dimensional ambient space while preserving the probability of failure. It is intended as a controlled numerical device for studying algorithmic behaviour in higher dimensions, rather than as a physically derived high-dimensional model. For this reason, these examples are complemented below by application examples that are intrinsically high-dimensional. Note that if the input distribution is a standard normal multivariate distribution, each of the $z_i$ will have a standard normal distribution.

For the purpose of reproducibility, the Python code and data used in this section are available on GitHub at \url{https://github.com/HughKinnear/nis_paper}. The \ac{SIS}, \ac{iCE} and \ac{EM} implementations were taken from \url{https://github.com/ERA-Software/Overview}. The reference probabilities for each numerical example were estimated with a \ac{MC} estimator, and are presented alongside the results tables.

\subsection{Piecewise Linear Function}\label{sec:pwl}

Table \ref{table:pwl} presents the results of the first numerical example, which revisits the piecewise linear function defined in Equation \ref{eq:piecewise linear}. For the 2-dimensional case, all methods have a reasonable mean estimate for the probability of failure. However, \ac{SIS} has a very large \ac{CoV}, and while \ac{iCE} has a much lower \ac{CoV}, it requires more performance function evaluations than \ac{NIS}, which has an even lower \ac{CoV}. The 100 and 300-dimensional cases have similar results, apart from the \ac{iCE} mean estimate, which underestimates the probability of failure by two orders of magnitude. This is because in these cases, \ac{iCE} struggles to populate the neighbourhood of the design point in any runs at all.

\begin{table}[ht]
\centering
\begin{tabular}{cccccc}
    \hline
    Dimension & Method & Mean $P_F$ & CoV $P_F$ & Mean $g$ evals \\
    \hline
    2    & SIS & $4.16 \times 10^{-5}$ & 2.80 & $1.25 \times 10^{4}$ \\
         & iCE & $2.92 \times 10^{-5}$ & 0.35 & $2.09 \times 10^{4}$ \\
         & NIS & $3.05 \times 10^{-5}$ & 0.07 & $1.44 \times 10^{3}$ \\
    \hline
    100  & SIS & $2.95 \times 10^{-5}$ & 2.31 & $1.26 \times 10^{4}$ \\
         & iCE & $2.73 \times 10^{-7}$ & 0.24 & $1.64 \times 10^{4}$ \\
         & NIS & $3.06 \times 10^{-5}$ & 0.10 & $9.42 \times 10^{3}$ \\
    \hline
    300  & SIS & $2.83 \times 10^{-5}$ & 2.73 & $1.25 \times 10^{4}$ \\
         & iCE & $2.99 \times 10^{-7}$ & 1.09 & $3.37 \times 10^{4}$ \\
         & NIS & $3.07 \times 10^{-5}$ & 0.11 & $2.44 \times 10^{4}$
\end{tabular}
\caption{Numerical results for the piecewise linear function. Reference probability $3.18 \times 10^{-5}$ estimated by \ac{MC} estimator with $10^8$ samples. The \ac{iCE} algorithm used $1$ initial component.}
\label{table:pwl}
\end{table}

\subsection{Meatball Function}\label{sec:meatball}

The second example revisits the meatball function from Equation \ref{eq:meatball}, and the results are presented in Table \ref{table:meatball}. For 2 dimensions, \ac{SIS} yields an entirely degenerate estimator, that is, the mean estimate is a large underestimate and the \ac{CoV} is very large. Both \ac{iCE} and \ac{NIS} produce a good estimator with a low \ac{CoV}, but \ac{NIS} is able to do so with a lower number of performance function evaluations. For the higher dimensional experiments, the \ac{iCE} estimator also becomes degenerate, whilst \ac{NIS} still performs well. Perhaps surprisingly, in the $300$-dimensional case, \ac{SIS} begins to populate the neighbourhood of the design point more frequently and hence it is able to produce a good mean estimate. This is likely due to the fact that the process used to increase the dimension of the reliability problem changes how the Markov chains move through the input space.

\begin{table}[ht]
\centering
\begin{tabular}{cccccc}
    \hline
    Dimension & Method & Mean $P_F$ & CoV $P_F$ & Mean $g$ evals \\
    \hline
    2    & SIS & $8.01 \times 10^{-6}$ & 5.22 & $1.47 \times 10^{4}$ \\
         & iCE & $1.10 \times 10^{-5}$ & 0.14 & $3.14 \times 10^{4}$ \\
         & NIS & $1.09 \times 10^{-5}$ & 0.08 & $2.62 \times 10^{3}$ \\
    \hline
    100  & SIS & $8.49 \times 10^{-6}$ & 4.06 & $1.52 \times 10^{4}$ \\
         & iCE & $1.66 \times 10^{-6}$ & 9.95 & $6.00 \times 10^{4}$ \\
         & NIS & $1.09 \times 10^{-5}$ & 0.11 & $1.80 \times 10^{4}$ \\
    \hline
    300  & SIS & $1.13 \times 10^{-5}$ & 3.50 & $1.47 \times 10^{4}$ \\
         & iCE & $3.48 \times 10^{-6}$ & 9.93 & $4.34 \times 10^{4}$ \\
         & NIS & $1.10 \times 10^{-5}$ & 0.09 & $5.25 \times 10^{4}$
\end{tabular}
\caption{Numerical results for the meatball function. Reference probability  $1.12 \times 10^{-5}$ estimated by \ac{MC} estimator with $10^8$ samples. The \ac{iCE} algorithm used $4$ initial components.}
\label{table:meatball}
\end{table}

\subsection{Two-degree-of-freedom Mass Spring System}\label{sec:two_dof}

\begin{figure}
    \centering
    \includegraphics[scale=0.35]{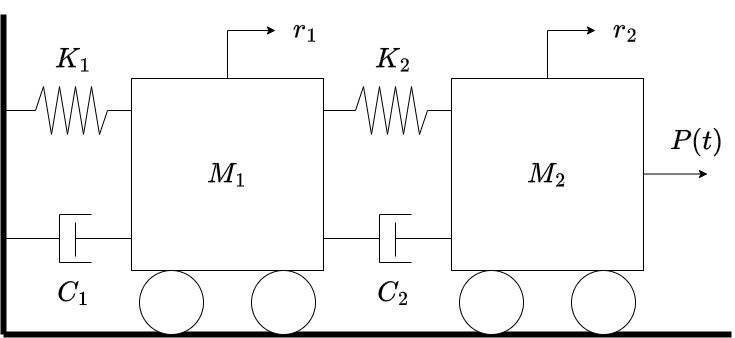}
    \caption{Two-degree-of-freedom mass spring system.}
    \label{fig:two_dof}
\end{figure}

The third example deals with the forced vibration of the \ac{TDOF} mass spring system \cite{sharmaModifiedReplicaExchangebased2023} depicted in Figure \ref{fig:two_dof}. The masses are $M_1 = M_2 = 2000 \text{ kg}$, the modal damping ratios are $\eta_1 = \eta_2 = 0.02$, and the forcing function acting on $M_2$ is given as $P(t)=2000\sin(11t)\text{N}$. Let the stiffness parameters $K_1$ and $K_2$ have independent log-normal distributions with mean $2.5 \times 10^5$ and \ac{CoV} 0.2. The standard normal input variables are transformed so that they have the required distribution, $K_1,K_2 = T(x_1,x_2)$. Let $r_1(t;K_1,K_2)$ be the displacement of the first mass at some time $t$, given stiffness constants. The performance function is
\begin{equation}\label{eq:two_dof}
    g_{\text{tdof}}(x_1,x_2) = \max_{0\leq t \leq20} r_1(t;K_1,K_2) - 0.024.
\end{equation}

This reliability problem has a failure region composed of two disconnected sets, where one of the sets is roughly three times larger than the other, with respect to the input distribution. For an accurate estimate of the probability of failure, both sets must be sampled. This is difficult for the \ac{SAIS} methods to accomplish on every run. However, \ac{NIS} is able to populate both sets at every run thanks to the \ac{NInitS} procedure.

Table \ref{table:tdof} shows the results of the numerical experiments. In the $2$-dimensional case, all three methods give a decent mean estimate for the probability of failure. However, \ac{SIS} has a large \ac{CoV}, and \ac{iCE} requires many more performance function evaluations than \ac{NIS}. In the higher dimensional cases, the \ac{CoV} of \ac{SIS} remains large whilst \ac{iCE} produces a biased estimate. \ac{NIS} is able to perform well even in high dimensions.

\begin{table}[ht]
\centering
\begin{tabular}{cccccc}
    \hline
    Dimension & Method & Mean $P_F$ & CoV $P_F$ & Mean $g$ evals \\
    \hline
    2    & SIS & $2.15 \times 10^{-5}$ & 2.62 & $1.00 \times 10^{4}$ \\
         & iCE & $1.87 \times 10^{-5}$ & 0.42 & $3.94 \times 10^{4}$ \\
         & NIS & $2.38 \times 10^{-5}$ & 0.08 & $1.86 \times 10^{3}$ \\
    \hline
    100  & SIS & $2.10 \times 10^{-5}$ & 3.10 & $1.00 \times 10^{4}$ \\
         & iCE & $7.94 \times 10^{-6}$ & 0.06 & $1.63 \times 10^{4}$ \\
         & NIS & $2.20 \times 10^{-5}$ & 0.12 & $1.87 \times 10^{4}$ \\
    \hline
    300  & SIS & $2.38 \times 10^{-5}$ & 3.54 & $1.01 \times 10^{4}$ \\
         & iCE & $6.92 \times 10^{-6}$ & 0.19 & $4.64 \times 10^{4}$ \\
         & NIS & $2.20 \times 10^{-5}$ & 0.11 & $4.66 \times 10^{4}$
\end{tabular}
\caption{Numerical results for the \ac{TDOF} performance function. Reference probability $2.48 \times 10^{-5}$ estimated by \ac{MC} estimator with $10^7$ samples. The \ac{iCE} algorithm used $2$ initial components.}
\label{table:tdof}
\end{table}

\subsection{Passive Vehicle Suspension Mechanical Model}\label{sec:suspension}

\begin{figure}
    \centering
    \includegraphics[scale=0.35]{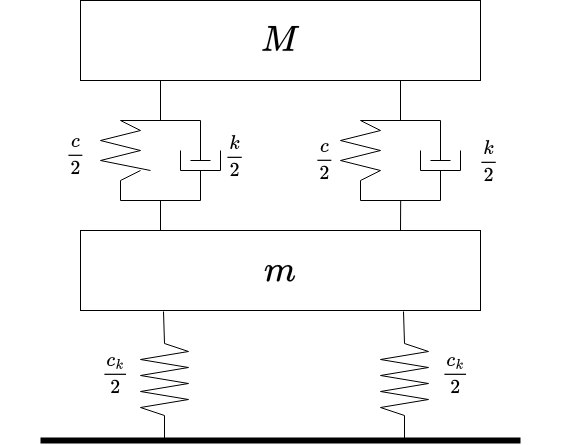}
    \caption{Passive vehicle suspension model.}
    \label{fig:suspension}
\end{figure}

This example considers the passive vehicle suspension model depicted in Figure \ref{fig:suspension}. The performance function, taken from \cite{rashkiSimulationbasedMethodReliability2014}, models the road-holding ability of a vehicle. The input variables are stiffness $c$ (kg/cm), tire stiffness $c_k$ (kg/cm) and damping coefficient $k$ (kg/cms), all with normal distributions with means and standard deviations of $(424,1480,47)$ and $(10,10,10)$ respectively. The input standard normal variables are transformed to have the correct distributions, $T(x_1,x_2,x_3) = (c,c_k,k)$, and the performance function is defined as
\begin{equation}\label{eq:suspension}
    g_{\text{veh}}(x_1,x_2,x_3) = 1- \left( \frac{\pi A V m}{b_0 G^2 k}\right)\left( \left( \frac{c_k}{M+m} -\frac{c}{M}\right)^2 + \frac{c^2}{Mm} + \frac{c_k k^2}{M^2m} \right),
\end{equation}
where $A=1 \text{ cm}^2/\text{cycle } m$, $b_0=0.27$, $V=10 \text{ m/s}$, $M = 3.2633 \text{ kg s}^2\text{/cm}$, $G = 981 \text{ cm/s}^2$ and $m=0.8158 \text{ kg s}^2\text{/cm}$.

This reliability problem has a topology similar to that of the piecewise linear function, but with more extreme gradient changes. The result is that \ac{SAIS} methods can frequently produce no failure samples at all. Table \ref{table:veh} shows the results of the numerical experiments and the reference probability of failure. In the 3-dimensional case \ac{SIS} does not produce any failure samples on any of the 100 experiment runs, whilst \ac{iCE} and \ac{NIS} both perform well, although \ac{NIS} uses much fewer performance function evaluations. For the 99 and 300-dimensional cases, the \ac{SIS} and \ac{iCE} estimators are both entirely degenerate, whereas \ac{NIS} is still able to perform well.

\begin{table}[ht]
\centering
\begin{tabular}{cccccc}
    \hline
    Dimension & Method & Mean $P_F$ & CoV $P_F$ & Mean $g$ evals \\
    \hline
    3    & SIS & $0                  $ & -    & $1.65 \times 10^{4}$ \\
         & iCE & $1.32 \times 10^{-6}$ & 0.25 & $1.14 \times 10^{4}$ \\
         & NIS & $1.42 \times 10^{-6}$ & 0.04 & $2.22 \times 10^{3}$ \\
    \hline
    99   & SIS & $0                  $ & -    & $1.67 \times 10^{4}$ \\
         & iCE & $1.25 \times 10^{-8}$ & 3.82 & $9.28 \times 10^{3}$ \\
         & NIS & $1.28 \times 10^{-6}$ & 0.10 & $1.17 \times 10^{4}$ \\
    \hline
    300  & SIS & $1.52 \times 10^{-8}$ & 6.52 & $1.64 \times 10^{4}$ \\
         & iCE & $5.36 \times 10^{-11}$ & 9.91 & $1.21 \times 10^{4}$ \\
         & NIS & $1.33 \times 10^{-6}$ & 0.12 & $2.84 \times 10^{4}$
\end{tabular}
\caption{Numerical results for the vehicle performance function. Reference probability $1.32 \times 10^{-6}$ estimated by \ac{MC} estimator with $10^8$ samples. The \ac{iCE} algorithm used $1$ initial component.}
\label{table:veh}
\end{table}

\subsection{Large Portfolio Losses}\label{sec:portfolio}

In this final example, a financial application \cite{elmasriImprovementCrossentropyMethod2021a} is considered. Note that this reliability problem does not have topology that is challenging for \ac{SIS} and \ac{iCE}. There is only one important niche. However, the niche itself has complex geometry, so the \ac{NInitS} procedure will often mistake the one niche for many, which means some performance function evaluations will be wasted. However, as \ac{NIS} iterates, it is often able to determine that there is only one effective niche, since the importance distribution it yields has many overlapping components. The example was chosen to demonstrate that \ac{NIS} performs well even under these circumstances.

Given a random variable $Z = (Z_1, \dots, Z_n)$, let the loss function be defined as
\begin{equation}\label{eq:loss function}
    L(Z) = \sum^n_{j=1} \one_{\{Z_j \geq 0.5 \sqrt{n}\}}.
\end{equation}
The failure region is defined as the loss function being greater than $bn$, where three cases are considered: $b=0.45, n=30$; $b=0.25, n=100$; $b=0.25, n=250$. The random variable $Z$ is defined as a function of other random variables,
\begin{equation}\label{eq:dependent}
    Z_j = \left(qU + (1-q^2)^{1/2}\eta_j \right) \mu^{-1/2}.
\end{equation}
where $U \sim \mathcal{N}(0,1)$, $\mu \sim \text{Gamma}(6,6)$, $\eta_j \sim \mathcal{N}(0,9)$ for $1 \leq j \leq n$, and $q=0.25$. The same realisation of $U$ and $\mu$ are used for all $Z_j$. The dimension of this reliability problem is $d = n+2$. These random variables must be transformed into standard normal space. That is, given standard normal input $\bm{x} =(x_1,x_2,\dots, x_d)$, $Z$ may be rewritten as,
\begin{equation}\label{eq:zj}
    Z_j = \left(qx_1 + 3(1-q^2)^{1/2}x_{j+2}\right)\left[ F^{-1}_\Gamma \left( F_{\mathcal{N}}(x_2)\right)\right]^{-1/2},
\end{equation}
for $1 \leq j \leq n$, where $F_{\mathcal{N}}$, $F_\Gamma$ are the cumulative distribution functions of $\mathcal{N}(0,1)$ and $\text{Gamma}(6,6)$ respectively. Finally, the portfolio loss performance function is defined,
\begin{equation}\label{eq:loss perf}
    g_{\text{loss}}(\bm{x}) = \sum^n_{j=1} \one_{\{Z_j \geq 0.5 \sqrt{n}\}} - bn.
\end{equation}
Note that a significant portion of the probability density of the failure region is represented by inputs which have a performance of exactly $0$. Since the inequality used in the definition of the probability of failure in this work was not strict, whilst it is in \cite{elmasriImprovementCrossentropyMethod2021a}, a small $\epsilon > 0$ is taken from the output of the performance functions in this numerical example.

Table \ref{table:loss} shows the results of the numerical experiments and the reference probability of failure. Most of the estimators for all of the dimensions perform well, though the \ac{SIS} estimator in the 252-dimensional case does have a larger \ac{CoV}. Despite the difficult geometry which deceives \ac{NInitS} into thinking there is more than one niche, \ac{NIS} has a comparable number of performance function evaluations and \ac{CoV} to the other methods for all the dimensions tested.

\begin{table}[ht]
\centering
\begin{tabular}{cccccc}
    \hline
    Dimension & Method & Mean $P_F$ & CoV $P_F$ & Mean $g$ evals \\
    \hline
    32   & SIS & $4.17 \times 10^{-3}$ & 0.21 & $7.86 \times 10^{3}$ \\
         & iCE & $4.24 \times 10^{-3}$ & 0.09 & $5.89 \times 10^{4}$ \\
         & NIS & $4.07 \times 10^{-3}$ & 0.09 & $1.04 \times 10^{4}$ \\
    \hline
    102  & SIS & $1.86 \times 10^{-3}$ & 0.22 & $8.00 \times 10^{3}$ \\
         & iCE & $1.83 \times 10^{-3}$ & 0.06 & $8.20 \times 10^{3}$ \\
         & NIS & $1.79 \times 10^{-3}$ & 0.12 & $1.43 \times 10^{4}$ \\
    \hline
    252  & SIS & $1.06 \times 10^{-5}$ & 1.45 & $1.04 \times 10^{4}$ \\
         & iCE & $1.08 \times 10^{-5}$ & 0.09 & $2.33 \times 10^{4}$ \\
         & NIS & $1.03 \times 10^{-5}$ & 0.08 & $3.56 \times 10^{4}$
\end{tabular}
\caption{Numerical results for the portfolio loss performance function. Reference probabilities are $4.28 \times 10^{-3}$, $1.81 \times 10^{-3}$ and $1.12 \times 10^{-5}$ for dimensions 32, 102 and 252 respectively. Reference probabilities estimated with \ac{MC} estimator with $10^7$ samples. The \ac{iCE} algorithm used $1$ initial component.}
\label{table:loss}
\end{table}

\ifSubfilesClassLoaded{%
    \newpage
    \bibliography{references}%
}

\end{document}

\section{Conclusions and Future Work}\label{sec:conclusion}

This paper proposes \ac{NIS}, a reliability method designed to robustly estimate the probability of failure for black-box problems that are high-dimensional and may exhibit challenging geometry. The development of the method was prompted by the idea of integrating hill valley testing, a niching technique taken from the field of evolutionary multi-modal optimisation, into an importance sampling framework. The numerical experiments suggest that its principal advantage arises on problems with multiple important niches or misleading geometric structure, whilst still showing strong performance on problems that do not exhibit the same challenges. The fundamental component of \ac{NIS} is \ac{NInitS}, an initial sampling procedure that is consistently able to populate all the important niches of a reliability problem.

The \ac{NInitS} procedure obviates the need for an initialisation routine in the \ac{EM} algorithm used to fit the mixture importance distribution. This is because the number of initial samples \ac{NInitS} returns provides a natural upper bound on the number of mixture components in the importance distribution. Reliability problems with multiple important niches are likely to cause ergodicity problems for any Markov chain algorithm. The result is that the \ac{EM} algorithm will yield a poor estimate for the mixture component weights when using these Markov chains to fit the parameters of a mixture distribution. A mixture component weight correction procedure was proposed, which is able to improve the estimate and specify chain weights that can help mitigate the problem on later iterations. A heuristic for determining the appropriate computational budget to assign to each Markov chain was also introduced, based on the mutual information of the mixture distribution.

All reliability methods have inevitable limitations that stem from their assumptions. Techniques such as \ac{iCE} and \ac{SIS} implicitly assume that the performance function does not have the types of challenging topology that have been examined in this work, and are able to perform well when this assumption is true. Conversely, \ac{NIS} does not assume this, and instead incurs a computational cost in order to understand the geometry of the failure region. In the case where the failure region comprises only one important niche, these extra performance evaluations will have been wasted. However, when the performance function is a black box, it is impossible to know this information a priori. Additionally, the computational cost of the \ac{NInitS} procedure is often relatively small compared to the computational cost of the rest of the \ac{NIS} algorithm.

It could be the case that a performance function has many important niches, and crucially, it could have more important niches than the user-defined maximum allowed initial samples. In this case, \ac{NIS} is almost guaranteed to underestimate the probability of failure. However, if once the algorithm has terminated, the estimated effective number of niches is very close to the maximum allowed initial samples, this could indicate to the user that there are more important niches to be modelled. In these circumstances, the entire \ac{NIS} algorithm could be run again with a higher number of maximum allowed initial samples. If a reliability problem has a very large number of important niches, explicitly modelling each individual niche may become more expensive than using a simple \ac{MC} estimator, especially when the probability of failure is relatively large. It is important to note that this type of reliability problem would present a serious and possibly insurmountable challenge for any variance reduction method.

Finally, note that whilst in this paper \ac{NInitS} has been developed for importance sampling, it is flexible enough so that in future work it could readily be utilised with other reliability methods, such as line sampling or surrogate modelling.

\newpage

\bibliography{references}

\end{document}